\documentstyle[12pt]{article}

\begin{document}

\title{The Free Energy Formula}

\author{Yury M. Zinoviev\thanks{This work is supported in part by the
Russian Foundation for Basic Research (Grant No. 96 - 01 - 00167)} \\
Steklov Mathematical Institute, \\
Gubkin St. 8, Moscow 117966, GSP - 1, Russia \\
e - mail: zinoviev@genesis.mi.ras.ru}

\date{}
\maketitle

\vskip 1cm

\noindent {\bf Abstract}. The Onsager formula for the free energy of the  
two dimensional Ising model with periodic boundary conditions is proved.

\vskip 1cm
\section{Introduction}

\noindent Onsager and Kaufman \cite{1}, \cite{2}, \cite{3} proposed the
formula for the partition function of the two dimensional Ising model.
Another heuristic method for the calculation of the partition function of
the two dimensional Ising model is proposed by Kac and Ward \cite{4}. 
Their idea consists of the construction of the special matrix $A$
whose determinant is connected with the partition function of the two
dimentional Ising model. They considered simultaneously two formulae:
the determinant of the matrix $A$ is proportional to the partition function
of the Ising model and it is proportional to the square of the partition
function. For the proof of the first formula they used a topological
statement. Sherman \cite{5} constructed a counter - example for this
statement. Hurst and Green \cite{6} proposed to use for the calculation 
of the Ising model partition function not a determinant but a Pfaffian of
some special matrix. This method was improved in the papers \cite{7}, 
\cite{8}. The proof proposed by McCoy and Wu \cite{9} is considered as
the most mathematically correct.

We consider a rectangular lattice on the plane formed by the points with
integral Cartesian coordinats
$x = j$, $y = k$, $0 \leq j \leq N$, $0 \leq k \leq M$, and the corresponding
horizontal and vertical edges in which the opposite sides of the entire 
rectangular are identified. In other words, we consider a rectangular lattice 
on a torus. McCoy and Wu \cite{9} proved the following formula for the
partition function of the Ising model
\begin{equation}
\label{1.1}
Z = (2\cosh \beta E_{1} \cosh \beta E_{2})^{MN}
1/2(- \hbox{Pf} \, \bar{A}_{1} + \hbox{Pf} \, \bar{A}_{2} + 
\hbox{Pf} \, \bar{A}_{3} + \hbox{Pf} \, \bar{A}_{4} ),
\end{equation}
where $\beta $ is the inverse temperature and $E_{1} (E_{2})$ is horizontal
(vertical) interaction energy. McCoy and Wu could not calculate the Pfaffians
$\hbox{Pf} \, \bar{A}_{i}$. They calculated the determinants 
$\det \bar{A}_{i}$ only. Since
\begin{equation}
\label{1.2}
\hbox{Pf} \, \bar{A}_{i}  = \pm (\det \bar{A}_{i})^{1/2},
\end{equation} 
it is necessary to calculate the sign in the formula (\ref {1.2}). 
These signs are calculated in \cite{9} by using the heuristic method.

In the paper \cite{10} a general formula of type (\ref {1.1}) is
obtained for the lattices placed on an arbitrary orientable surface.
In this paper we prove that for the torus case the general formula \cite{10}
is the following formula
\begin{equation}
\label{1.3}
Z = ( - 2\cosh \beta E_{1} \cosh \beta E_{2})^{MN}
1/2(- \hbox{Pf} \, \bar{A}_{1} + \hbox{Pf} \, \bar{A}_{2} + 
\hbox{Pf} \, \bar{A}_{3} + \hbox{Pf} \, \bar{A}_{4} ),
\end{equation}
We will calculate the Pfaffians $\hbox{Pf} \, \bar{A}_{i}$ and we will
prove the Onsager formula for the free energy of the two dimensional Ising 
model. It is easy to guess our formulae for the Pfaffians by using the 
formulae for $\det \bar{A}_{i}$ given in the book \cite{9} and the equality
\begin{eqnarray}
\label{1.4}
(1 + z_{1}^{2})(1 + z_{2}^{2}) - 2z_{1}(1 - z_{2}^{2}) - 
2z_{2}(1 - z_{1}^{2}) = \nonumber \\ 
(z_{1}z_{2} + z_{1} + z_{2} - 1)^{2}.
\end{eqnarray}

\section{Homological Formula}
\setcounter{equation}{0}

\noindent We denote by $G(N,M)$ the graph described in the previuos section.
Let the function $\sigma$ on the vertices of the graph $G(N,M)$ takes the
values in the multiplicative group ${\bf Z}_{2} = \{1,- 1\}$.  The energy
$H(\sigma )$ for the Ising model with zero magnetic field can be expressed
in the form
\begin{equation}
\label{2.1}
H(\sigma ) = - \sum_{j = 1}^{N} \sum_{k = 1}^{M}
(E_{1}\sigma (j,k)\sigma (j + 1,k) + E_{2}\sigma (j,k)\sigma (j,k + 1)),
\end{equation}
where the numbers $1$ and $N + 1$ are identified and the numbers $1$  and
$M + 1$ are identified also. The partition function of the Ising model
with zero magnetic field is defined as follows
\begin{equation}
\label{2.2}
Z(N,M) = \sum_{\sigma} \exp \{ - \beta H(\sigma)\},
\end{equation}
where the summing runs over the group of the functions $\sigma$  on the
vertices of the graph $G(N,M)$ taking values in the group ${\bf Z}_{2}$. 
By performing the Fourier transformation \cite{11}, (\cite{9}, Chapter 5) 
we have
\begin{equation}
\label{2.3}
Z(N,M) = (2\cosh\beta E_{1}\cosh\beta E_{2})^{MN}
\sum_{\xi} \prod_{i = 1}^{2} (\tanh \beta E_{i})^{k_{i}(\xi)},
\end{equation} 
The function $\xi$ on the edges of the graph $G(N,M)$ is called the cycle
with the coefficients in the group ${\bf Z}_{2}$ if the product of all
values of the function $\xi$ on the edges connecting an arbitrary vertex of
$G(N,M)$ to other vertices is equal to $1$. The number 
$k_{1}(\xi) (k_{2}(\xi))$ is the total number of all horizontal 
(vertical) edges on which the cycle $\xi$ takes the value $- 1$.

By a dimer configuration on the graph $G$ we mean a system of edges
$U = \{ (p_{k},q_{k})\}$ satisfying the following conditions:  
the edges of $U$ are pairwise disjoint; all vertices of the graph $G$ 
are covered by the edges of the dimer configuration $U$. 

Due to \cite{10} we will establish a one - to - one correspondence between
the elements of the group of cycles with coefficiens in ${\bf Z}_{2}$ and
the dimer configurations on another, expanded graph. Every vertex of the
initial graph $G(N,M)$ is connected with four other vertices. In the new
graph  we replace every vertex with two new vertices and add to the old
four edges one new edge connecting two new vertices. We assume that new
vertices are the end points of two old edges and one new edge. Thus every
vertex is connected with exactly three other vertices. By applying this
construction for every vertex of the graph $G(N,M)$ we obtain a new graph
$tr(G(N,M))$. Let a cycle $\xi$ be given on edges of the graph $G(N,M)$.
Then it is given on the old edges of the graph $tr(G(N,M))$. By using the
condition $\partial \hat{\xi} = - 1$ we continue this function $\xi$ as
${\bf Z}_{2}$ valued function $\hat{\xi}$ on the edges of the graph
$tr(G(N,M))$. Namely for every vertex of the graph $tr(G(N,M))$ the product
of the function $\hat{\xi}$ values on the edges connecting this vertex
with three other vertices of the graph $tr(G(N,M))$ is equal to $-1$.
The function $\xi$ is a cycle and every vertex of the graph $G(N,M)$ 
corresponds with two vertices of the graph $tr(G(N,M))$. Therefore the
extension $\hat{\xi}$ on the edges of the graph $tr(G(N,M))$ is defined
uniquely. Conversely the restriction of every function $\hat{\xi}$ on the
edges of the graph $tr(G(N,M))$, satisfying the condition
$\partial \hat{\xi} = - 1$, to the edges of the graph $G(N,M)$ defines a 
cycle on the edges of the graph $G(N,M)$ with coefficients in the group 
${\bf Z}_{2}$.

We change every vertex of the graph $tr(G(N,M))$ by a triangle each vertex 
of which is an end point of one edge of the graph $tr(G(N,M))$ and an
end point of the two edges of triangle. We obtain a new graph $cl(G(N,M))$.
The set of the functions $\hat{\xi}$ on the edges of the graph $tr(G(N,M))$,
satisfying condition $\partial \hat{\xi} = - 1$, is in one - to - one 
correspondence with the set of all dimer configurations on the graph
$cl(G(N,M))$. Indeed, let us include into the dimer configuration 
$U(\hat{\xi})$ all the edges on which the function $\hat{\xi}$ takes the 
value $- 1$. Then three or one such edges come to every triangle. In the
first case all vertices of the triangle are covered by the edges which
do not intersect each other and some edges of the triangle does not belong to 
the dimer configuration $U(\hat{\xi})$. In the second case one edge of 
the dimer configuraton $U(\hat{\xi})$ comes to one vertex of the triangle,
and we include into the dimer configuration the opposite side of the 
triangle. Now again all vertices of the triangle are covered by the edges 
from the dimer configuration $U(\hat{\xi})$ which do not intersect each
other.

Let the dimer configuration $U$ on the graph $cl(G(N,M))$ be given.
Let us define the function $\hat{\xi} (U)$ by taking it is equal to
$1$ on the edges of the graph $tr(G(N,M))$ which do not belong to the
dimer configuration $U$ and by taking it equal to $- 1$ on the edges of 
the graph $tr(G(N,M))$ which belong to the dimer configuration $U$. 
If one side of a triangle belongs to the dimer configuration $U$, then
one edge from the dimer configuration $U$ comes to a triangle. If the sides
of a triangle do not belong to the dimer configuration $U$, then three edges 
from the dimer configuration $U$ come to a triangle. In both cases the
function $\hat{\xi}(U)$ satisfies the condition $\partial \hat{\xi} = - 1$.

One - to - one correspondence allows us to rewrite the partition function 
(\ref{2.3}) in another form. Every dimer configuration $U$ on the graph
$cl(G(N,M))$ corresponds with a cycle $\xi$  on the edges of the graph 
$G(N,M)$ with coefficients in the group ${\bf Z}_{2}$: 
$(p,q) \in U\cap G(N,M)$ if and only if $\xi((p,q)) = - 1$. Therefore the
relation (\ref{2.3}) can be rewritten in the following form
\begin{equation}
\label{2.4}
Z(N,M) = (2\cosh \beta E_{1}\cosh \beta E_{2})^{MN}
\sum_{U} \prod_{i = 1}^{2} (\tanh \beta E_{i})^{\# (U\cap G(N,M))_{i}}, 
\end{equation}
where summation runs over all dimer configurations $U$ on the graph
$cl(G(N,M))$. $\# (U\cap G(N,M))_{1}$ ($\# (U\cap G(N,M))_{2}$) 
denotes the total number of horizontal (vertical) common edges in the dimer
configuration $U$ and in the graph $G(N,M)$.

With each vertex of the graph $G(N,M)$ there corresponds a cluster of the
graph 

\noindent
$cl(G(N,M))$: two triagles connected by one edge. We place one 
triangle under another such that their bases are horizontal. Let vertex
opposite to the base of the upper triangle be set below the base and the
vertex opposite to the base of the lower triangle be set above the base.
The vertical edge connects these two vertices. The right vertex of the 
upper triangle is denoted by $1$. The left vertex of the lower triangle 
is denoted by $2$. The left vertex of the upper triangle is denoted by $3$.
The right vertex of the lower triangle is denoted by $4$. The upper vertex 
of the lower triangle is denoted by $5$. The lower vertex of the upper 
triangle is denoted by $6$. With all six vertices there correspond two 
numbers $(j,k)$, $1 \leq j \leq N$, $1 \leq k \leq M$ corresponding to the
old vertex of the graph $G(N,M)$. The vertices
$(1,j,k)$ and $(6,j,k)$; $(1,j,k)$ and $(3,j,k)$; $(3,j,k)$ and $(6,j,k)$;
$(5,j,k)$ and $(6,j,k)$; $(2,j,k)$ and $(4,j,k)$; $(2,j,k)$ and $(5,j,k)$;
$(4,j,k)$ and $(5,j,k)$ are connected by the edges of one cluster. The
vertices $(1,j,k)$ and $(2,j + 1,k)$ are connected by the horizontal edges
of the graph $G(N,M)$. We consider the numbers $1$ and $N + 1$ identical.
The vertices $(3,j,k)$ and $(4,j,k + 1)$ are connected by the vertical edges
of the graph $G(N,M)$. The numbers $1$ and $M + 1$ are considered identical.

Let an orientation of every edge of the graph $cl(G(N,M))$ be given. If an
edge is defined by its end points $(p,q)$, then an orientation is given by
a function $\phi (p,q)$ taking only two values $0$, $1$ and satisfying the
following condition
\begin{equation}
\label{2.5}
\phi (p,q) + \phi (q,p) = 1 \; \hbox{mod} \, 2.
\end{equation} 

Let us define the function $z(p,q) = z(q,p)$ on the oriented edges $(p,q)$: 
$z(p,q) = 1$ if the vertices $p,q$ belong to the same cluster  and are
connected by an edge; $z(p,q) = \tanh \beta E_{1}$ 
($z(p,q) = \tanh \beta E_{2}$) if the vertices $p,q$ belong to the
neighboring clusters and are connected by a horizontal (vertical) edge;
$z(p,q) = 0$ if the vertices $p,q$ are not connected by an edge.

Let the linear ordering of vertices of the graph $cl(G(N,M))$ be given
\begin{equation}
\label{2.6}
n(i,j,k) = i + 6(k - 1) + 6M(j - 1)
\end{equation}
where $1 \leq i \leq 6$, $1 \leq j \leq N$, $1 \leq k \leq M$. The function
(\ref{2.6}) establishes the correspondence between the vertices of the graph 
$cl(G(N,M))$ and the numbers $1,...,6MN$.

Let us define $6MN \times 6MN$ - matrix
\begin{equation}
\label{2.7}
A(\phi)_{n(p)n(q)} = \exp \{ i\pi \phi (p,q)\}z(p,q)
\end{equation}
where the numbers $p = (i,j,k)$ are indices for the vertices of the graph
$cl(G(N,M))$. Due to the relation (\ref{2.5}) the matrix (\ref{2.7}) is
antisymmetric. Hence we can define its Pfaffian
\begin{equation}
\label{2.8}
\hbox{Pf} \, A(\phi) = ((3MN)!)^{- 1}2^{- 3MN} \sum_{\pi \in S_{6MN}}
(- 1)^{\sigma (\pi )} A(\phi )_{\pi (1)\pi (2)} \cdots
A(\phi )_{\pi (6MN - 1)\pi (6MN)}
\end{equation}
where $\pi$ is an arbitrary permutation of the numbers $1,...,6MN$ and
$\sigma (\pi )$ is its parity.

The definition (\ref{2.8}) can be expressed as
\begin{equation}
\label{2.9}
\hbox{Pf} \, A(\phi ) = \sum_{U} (- 1)^{\sigma (U)} \prod_{(i,j) \in U}
A(\phi )_{i,j}
\end{equation}
where $U$ is an ordered subdivision of the numbers $1,...,6MN$ into pairs

\noindent
$\{ (i_{1},j_{2}),...,(i_{3MN},j_{3MN})\}$,  
$1 = i_{1} < \cdots < i_{3MN} \leq 6MN$, $i_{p} < j_{p}$,
$p = 1,...,3MN$. The number $\sigma (U) = 0,1$ respectively with the parity 
of the permutation mapping the numbers $(i_{1},j_{1},...,i_{3MN},j_{3MN})$
into the numbers $(1,...,6MN)$. Due to the relation (\ref{2.7}) and the
definition of the function $z(p,q)$ the summation in (\ref{2.9}) runs 
only such subdivitions $U$ which correspond with the dimer configurations
on the graph $cl(G(N,M))$. Let $U = \{ (p_{1},q_{1}),...,(p_{3MN},q_{3MN})\}$ 
be a dimer configuration on the graph $cl(G(N,M))$. By using an orientation
function $\phi$ satisfying the condition (\ref{2.5}) we define a function on 
the dimer configurations on the graph $cl(G(N,M))$
\begin{equation}
\label{2.10}
\tau [\phi ](U) = \sigma (n(p_{1}),n(q_{1}),...,n(p_{3MN}),n(q_{3MN})) +
\sum_{j = 1}^{3MN} \phi (p_{j},q_{j}) \; \hbox{mod} \, 2   
\end{equation} 
where $\sigma (n(p_{1}),n(q_{1}),...,n(p_{3MN}),n(q_{3MN})) = 0,1$
respectively with the parity of the permutation mapping the numbers
$(n(p_{1}),n(q_{1}),...,n(p_{3MN}),n(q_{3MN}))$ into the numbers $(1,...,6MN)$.
The function $\phi (p,q)$ satisfies the condition (\ref{2.5}). Hence the
the right hand side of the equality (\ref{2.10}) is independent of an 
ordering of the vertices in the edges $(p_{j},q_{j})$ although every
summand in (\ref{2.10}) depends on this ordering. The right hand side of
the equality (\ref{2.10}) is also independent of an ordering of the edges
$(p_{j},q_{j})$. 

The substitution of the definitions (\ref{2.7}) and (\ref{2.10}) into 
the right hand side of the equality (\ref{2.9}) yields
\begin{equation}
\label{2.11}
\hbox{Pf} \, A(\phi ) = \sum_{U} (- 1)^{\tau [\phi ](U)} z(U) 
\end{equation}
where
\begin{equation}
\label{2.12}
z(U) = \prod_{(p,q) \in U} z(p,q)
\end{equation}
and the summation in (\ref{2.11}) runs over all dimer configurations on the
graph $cl(G(N,M))$.  

The expression (\ref{2.11}) differs from the sum (\ref{2.4}) by the 
multiplier $(- 1)^{\tau [\phi ](U)}$ only. For any dimer configuration
$U_{0}$ on the graph $cl(G(N,M))$ we have
\begin{equation}
\label{2.13}
\hbox{Pf} \, A(\phi ) = (- 1)^{\tau [\phi ](U_{0})} \sum_{U} 
(- 1)^{\tau [\phi](U_{0}) + \tau [\phi](U)} z(U).
\end{equation}

Let us study the function $\tau [\phi ](U_{1}) + \tau [\phi ](U_{2})$ 
of a pair of the dimer configurations $U_{1}$, $U_{2}$ on the graph 
$cl(G(N,M))$. Let us consider the symmetric difference
\begin{equation}
\label{2.14}
U_{1}\triangle U_{2} = (U_{1}\cup U_{2}) \setminus (U_{1}\cap U_{2}).
\end{equation}
The dimer configuration definition implies that this set consists of the
finite number of closed broken lines which do not intersect each other,
that is, we can write
\begin{equation}
\label{2.15}
U_{1}\triangle U_{2} = \cup_{i = 1}^{s} \gamma_{i}.
\end{equation}
Every broken line $\gamma_{i}$ does not intersect itself. The edges from
$U_{1}$ and $U_{2}$ are included in each broken line alternatively. 
Due to Theorem 2 from \cite{10} for any dimer configurations $U_{1}$,  
$U_{2}$ on the graph $cl(G(N,M))$
\begin{equation}
\label{2.16}
\tau [\phi ](U_{1}) + \tau [\phi ](U_{2}) = s + \sum_{i = 1}^{s}
\phi (\gamma_{i} ) \; \hbox{mod} \,2 
\end{equation}
where the broken lines $\gamma_{i} $  are given by the relation (\ref{2.15}) 
and
\begin{equation}
\label{2.17}
\phi (\gamma_{i} ) = \sum_{(p,q) \in \gamma_{i}} \phi (p,q) \; \hbox{mod} \, 2.
\end{equation}
This sum does not depend on the orietation chosen on the broken line 
$\gamma_{i}$ since the total number of edges in the broken line
$\gamma_{i}$ is even.

Let us define on the graph $cl(G(N,M))$ special orientation function 
\begin{eqnarray}
\label{2.18}
\phi ((1,j,k),(6,j,k)) = \phi ((6,j,k),(3,j,k)) = \nonumber \\
\phi ((3,j,k),(1,j,k)) = 0 \; \hbox{mod} \, 2 \nonumber \\
\phi ((2,j,k),(5,j,k)) = \phi ((5,j,k),(4,j,k)) = \nonumber \\
\phi ((4,j,k),(2,j,k)) = 0 \; \hbox{mod} \, 2 \nonumber \\
\phi ((5,j,k),(6,j,k)) = 0 \; \hbox{mod} \, 2 \nonumber \\
\phi ((1,j,k),(2,j + 1,k)) = 0 \; \hbox{mod} \, 2 \nonumber \\
\phi ((3,j,k),(4,j,k + 1)) = 0 \; \hbox{mod} \, 2.
\end{eqnarray}
In the equalities (\ref{2.18}) the numbers $1$  and $N + 1$, $1$ and $M + 1$
are identified. The graph $cl(G(N,M))$ consists of the triangles and $12$ -
angles. The relations (\ref{2.18}) have the property: if a broken line 
$\gamma$  is a boundary of a triangle or $12$ - angle and the counter -
clockwise orientation is chosen, then
\begin{equation}
\label{2.19}
\phi (\gamma) = 1 \; \hbox{mod} \, 2 .
\end{equation}
{\bf Theorem 2.1} {\it Let on the graph} $cl(G(N,M))$ {\it the orientation
function} (\ref{2.18}) {\it be given. Let domain on the torus be bounded
by the broken lines} $\gamma_{1},...,\gamma_{s}$ {\it and these broken lines
coincide with the symmetric difference of two dimer configurations on the
graph} $cl(G(N,M))$. {\it Then}
\begin{equation}
\label{2.20}
s + \sum_{i = 1}^{s} \phi (\gamma_{i}) = 0 \; \hbox{mod} \, 2.
\end{equation}
{\it Proof.} In view of the condition of the theorem the domain on the torus
is bounded by broken lines $\gamma_{1},...,\gamma_{s}$ which do not 
intersect each other. We stretch the auxiliary disk $d_{i}$ on each broken
line $\gamma_{i}$. In general, we obtain the new closed orientable surface 
$P$. The domain consists of the faces $\sigma$ (triangles and $12$ - angles).
We choose the orientation which induces the counter- clockwise orientation
on the broken lines $\partial \sigma$. Let us choose an orientation of
each auxillary disk $d_{i}$ coherently with the orientation of the 
boundary of the domain.

Since the set of closed broken lines $\gamma_{1},...,\gamma_{s}$  coicide
with the symmetric difference of two dimer configurations, the domain
contains an even number of vertices, the number of edges and the number 
of vertices for each broken line $\gamma_{i}$ are also even. Therefore
the number of vertices $v$ of the constructed cell complex for an orietable
closed surface $P$ is even. Let $f$  and $e$ denote the total numbers of
faces and edges of the orientable surface $P$. By using the skew symmetry 
condition (\ref{2.5}) and by summing over all faces of the surface $P$
including the auxiliary disks $d_{i}$ we have
\begin{equation}
\label{2.21}
\sum_{\sigma \in P} (1 + \phi (\partial \sigma)) = f + e = v + \chi = 0
\; \hbox{mod} \, 2,
\end{equation} 
since Euler characteristic $\chi = 2(1 - g)$ of the orientable surface
$P$ is even.

If the orientation function $\phi$ on the graph $cl(G(N,M))$ is given by
the relations (\ref{2.18}), then for any face of the surface $P$ excepting
the auxiliary disks the relation (\ref{2.19}) holds. Hence the left hand
sides of the  equalities (\ref{2.20}) and (\ref{2.21}) coincide. The
theorem is proved.

\noindent {\bf Theorem 2.2} {\it Let the dimer configuration} $U_{0}$ 
{\it on the graph} $cl(G(N,M))$ {\it correspond to the cycle} $\xi_{0}$ 
{\it with the coefficients in the multiplicative group} ${\bf Z}_{2}$
{\it which takes the value} $1$ {\it on all edges of the graph} $G(N,M)$. 
{\it Let the function} $\tau [\phi ](U)$ {\it be given by the relation} 
(\ref{2.10}) {\it for the orietation function} $\phi$ {\it defined by the
equalities} (\ref{2.18}). {\it Then}
\begin{equation}
\label{2.22}
\tau [\phi ](U_{0}) = MN \; \hbox{mod} \, 2.
\end{equation}
{\it Proof.} The dimer configuration $U_{0}$ does not contain the edges
connecting different clusters. In every cluster $(j,k)$ the dimer 
configuration $U_{0}$ contains three edges:
$((1,j,k),(3,j,k))$, $((2,j,k),(4,j,k))$ and $((5,j,k),(6,j,k))$. In
according to (\ref{2.6})
\begin{eqnarray}
\label{2.23}
\sigma (n(1,1,1),n(3,1,1),n(2,1,1),n(4,1,1),n(5,1,1),n(6,1,1),..., \nonumber \\
n(1,N,M),n(3,N,M),n(2,N,M),n(4,N,M),n(5,N,M),n(6,N,M)) = \nonumber \\
MN \; \hbox{mod} \, 2
\end{eqnarray}
since there is one disorder: $n(3,j,k),n(2,j,k)$ in any of $MN$ clusters.

The definitions (\ref{2.18}) and the relation (\ref{2.5}) imply
\begin{equation}
\label{2.24}
\sum_{j = 1}^{N} \sum_{k = 1}^{M} [\phi ((1,j,k),(3,j,k)) +
\phi ((2,j,k),(4,j,k)) + \phi ((5,j,k),(6,j,k))] = 0 \; \hbox{mod} \, 2.
\end{equation}
The substitution of the equalities (\ref{2.23}) and (\ref{2.24}) into the 
equality (\ref{2.10}) for the dimer configuration $U_{0}$ yields the 
equality (\ref{2.22}). The theorem is proved.

The number (\ref{2.22}) is the total number of the vertices of the graph
$G(N,M)$ or it is the total number of the clusters of the graph
$tr(G(N,M))$. The formulae (\ref{1.1}) and (\ref{1.3}) differ each other
by the multiplier $(- 1)^{\tau [\phi ](U_{0})}$.

We denote by ${\bf Z}_{2}^{add}$ the group of modulo $2$ residuals. The
modulo $2$ residuals are multiplied each other and the group 
${\bf Z}_{2}^{add}$ is a field. The graph $cl(G(N,M))$ divides the torus
into the faces: the triangles and $12$ - angles. The cell complex 
$P(cl(G(N,M))) \equiv P(G)$ is called the set consisting of the cells
(vertices, edges, faces). To every cell $s_{i}^{p}$ there corresponds the
natural number $p$ (dimension). For the vertices $p = 0$, for the edges
$p = 1$ and for the faces $p = 2$. To every pair of the cells
$s_{i}^{p}$, $s_{j}^{p - 1}$ there corresponds the number
$(s_{i}^{p}:s_{j}^{p - 1}) \in {\bf Z}_{2}^{add}$ (incidence number).
If the cell $s_{j}^{p - 1}$ is included into the boundary of the cell
$s_{i}^{p}$, then the incidence number $(s_{i}^{p}:s_{j}^{p - 1}) = 1$. 
Otherwise the incidence number $(s_{i}^{p}:s_{j}^{p - 1}) = 0$. 
For any pair of the cells $s_{i}^{2}$, $s_{j}^{0}$ the incidence numbers
satisfy the condition
\begin{equation} 
\label{2.25}
\sum_{m} (s_{i}^{2}:s_{m}^{1})(s_{m}^{1}:s_{j}^{0}) = 0 \; \hbox{mod} \, 2.
\end{equation}
Indeed, if the vertex $s_{j}^{0}$ is not contained in the boundary of the
triangle or
$12$ - angle $s_{i}^{2}$, then the condition (\ref{2.25}) is fulfilled. If the
vertex $s_{j}^{0}$  is included into the boundary of the face $s_{i}^{2}$,
then it is included into the boundaries of three edges $s_{m}^{1}$ two of 
which are included into the boundary of the face $s_{i}^{2}$. The condition 
(\ref{2.25}) is fulfilled again.

A cochain $c^{p}$ of the complex $P(G)$ with the coefficients in the group
${\bf Z}_{2}^{add}$ is a function on the $p$ dimensional cells taking 
values into the group ${\bf Z}_{2}^{add}$. Usually the cell orientation
is considered and the cochains are the antisymmetric functions:
$c^{p}(- s^{p}) = - c^{p}(s^{p})$. However, $ - 1 = 1 \; \hbox{mod} \, 2$ 
and we can neglect the cell orientation for the coefficients in the group
${\bf Z}_{2}^{add}$. The cochains form an Abelian group
\begin{equation}
\label{2.26}
(c^{p} + c^{\prime p})(s_{i}^{p}) = c^{p}(s_{i}^{p}) + 
c^{\prime p}(s_{i}^{p}) \; \hbox{mod} \, 2.
\end{equation}  
It is denoted by $C^{p}(P(G),{\bf Z}_{2}^{add})$. The mapping
\begin{equation}
\label{2.27}
\partial c^{p}(s_{i}^{p - 1}) = \sum_{j} (s_{j}^{p}:s_{i}^{p - 1})
c(s_{j}^{p}) \; \hbox{mod} \, 2
\end{equation}
defines the homomorphism of the group $C^{p}(P(G),{\bf Z}_{2}^{add})$ into
the group $C^{p - 1}(P(G),{\bf Z}_{2}^{add})$. It is called the 
boundary operator. The mapping
\begin{equation}
\label{2.28}
\partial^{\ast} c^{p}(s_{i}^{p + 1}) = \sum_{j} (s_{i}^{p + 1}:s_{j}^{p})
c(s_{j}^{p}) \; \hbox{mod} \, 2
\end{equation}
defines the homomorphism of the group $C^{p}(P(G),{\bf Z}_{2}^{add})$ into
the group $C^{p + 1}(P(G),{\bf Z}_{2}^{add})$. It is called the cobounbary 
operator. The condition (\ref{2.25}) implies $\partial \partial = 0$, 
$\partial^{\ast} \partial^{\ast} = 0$. A kernel 
$Z_{p}(P(G),{\bf Z}_{2}^{add})$ of the homomorphism
$\partial : C^{p}(P(G),{\bf Z}_{2}^{add}) \rightarrow
C^{p - 1}(P(G),{\bf Z}_{2}^{add})$ is called a group of cycles of complex
$P(G)$ with the coefficients in the group ${\bf Z}_{2}^{add}$. The
image $B_{p}(P(G),{\bf Z}_{2}^{add})$ of the homomorphism
$\partial : C^{p + 1}(P(G),{\bf Z}_{2}^{add}) \rightarrow
C^{p}(P(G),{\bf Z}_{2}^{add})$ is called a group of boundaries of the 
complex $P(G)$ with the coefficients in the group ${\bf Z}_{2}^{add}$. 
Since $\partial \partial = 0$, a group $B_{p}(P(G),{\bf Z}_{2}^{add})$
is a subgroup of the group $Z_{p}(P(G),{\bf Z}_{2}^{add})$.
A quotient group $Z_{p}(P(G),{\bf Z}_{2}^{add})/
B_{p}(P(G),{\bf Z}_{2}^{add})$ is called a homology group 
$H_{p}(P(G),{\bf Z}_{2}^{add})$ of the complex $P(G)$ with the coefficients
in the group ${\bf Z}_{2}^{add}$. Similarly, for the coboundary operator
$\partial^{\ast}$ a group of cocycles $Z^{p}(P(G),{\bf Z}_{2}^{add})$, 
a group of coboundaries $B^{p}(P(G),{\bf Z}_{2}^{add})$ and a group of
cohomologies $H^{p}(P(G),{\bf Z}_{2}^{add})$ are defined.

It is possible to introduce the bilinear form on 
$C^{p}(P(G),{\bf Z}_{2}^{add})$ 
\begin{equation}
\label{2.29}
\langle f^{p},g^{p} \rangle = \sum_{i} f^{p}(s_{i}^{p})g^{p}(s_{i}^{p})
\; \hbox{mod} \, 2.
\end{equation}
The definitions (\ref{2.27}) and (\ref{2.28}) imply
\begin{equation}
\label{2.30}
\langle f^{p},\partial^{\ast} g^{p - 1}\rangle =
\langle \partial f^{p},g^{p - 1}\rangle \; \hbox{mod} \, 2
\end{equation}
\begin{equation}
\label{2.31}
\langle f^{p},\partial g^{p + 1}\rangle =
\langle \partial^{\ast} f^{p},g^{p + 1}\rangle \; \hbox{mod} \, 2.
\end{equation}

We identify every set of cells $\{ s_{i}^{p}\}$ with a cochain taking value
$1$ on these cells and which takes value $0$ on other cells. For example,
the symmetric difference (\ref{2.15}) of two dimer configurations
$U_{1}$, $U_{2}$ on the graph $cl(G(N,M))$ is identified with the cochain
which takes the value $1$ on the edges belonging to the broken lines 
$\gamma_{i}$ and which takes the value $0$ on the other edges of the graph
$cl(G(N,M))$.

In according to the relation (\ref{2.13}) it is necessary to study the
function
\begin{equation}
\label{2.32}
\psi (U\triangle U_{0}) = \tau [\phi ](U) + \tau [\phi ](U_{0}) \; \hbox{mod} 
\, 2
\end{equation}
where the function $\tau [\phi ](U)$ is defined by the relation (\ref{2.10})
by using the orientation function (\ref{2.18}). For the dimer configuration
$U_{0}$ we take one corresponding to the cycle $\xi_{0}$ with the coefficients 
in the group ${\bf Z}_{2}$ which takes value $1$ on all edges of the graph
$G(N,M)$. (We consider on the graph $G(N,M)$ the cycles with the coefficients 
in the multiplicative group ${\bf Z}_{2}$. We hope that making use of 
the cycles with the coefficients in the different groups ${\bf Z}_{2}$ and 
${\bf Z}_{2}^{add}$ on the different graphs does not put us to cofusion.)
Due to relations (\ref{2.15}), (\ref{2.16})
\begin{equation}
\label{2.33}
U\triangle U_{0} = \sum_{i = 1}^{s} \gamma_{i}
\end{equation}
\begin{equation}
\label{2.34}
\psi (U\triangle U_{0}) = s + \sum_{i = 1}^{s} \gamma_{i} \; \hbox{mod} \, 2.
\end{equation}
{\bf Theorem 2.3} {\it The function} $\psi (U\triangle U_{0})$,{\it given
by the relations} (\ref{2.32}) - (\ref{2.34}), {\it depends on the 
equivalence class of the cycle} $U\triangle U_{0}$ {\it in the homology
group} $H_{1}(P(G),{\bf Z}_{2}^{add})$ {\it only}.

\noindent {\it Proof.}  Due to relation (\ref{2.33}) the symmetric 
difference $U\triangle U_{0}$ is the set of the broken closed lines. 
Hence it is identified with the cycle from the group 
$Z_{1}(P(G),{\bf Z}_{2}^{add})$. Thus the symmetric difference
$U\triangle U_{0}$ generates some equivalence class in the homology group 
$H_{1}(P(G),{\bf Z}_{2}^{add})$. Let for the dimer configurations
$U_{1}$ and $U_{2}$ on the graph $cl(G(N,M))$ the symmetric differences
$U_{1}\triangle U_{0}$ and $U_{2}\triangle U_{0}$ generate the same 
equivalence class in the homology group. Let us prove
$\psi (U_{1}\triangle U_{0}) = \psi (U_{2}\triangle U_{0})$. Indeed
in this case the symmetric difference
$(U_{1}\triangle U_{0})\triangle (U_{2}\triangle U_{0})$ is a boundary of
some domain on the torus. Due to Lemma 1 from \cite{10}
\begin{equation}
\label{2.35}
(U_{1}\triangle U_{0})\triangle (U_{2}\triangle U_{0}) =
U_{1}\triangle U_{2}.
\end{equation} 
Therefore the symmetric difference $U_{1}\triangle U_{2}$ is a boundary
of some domain on the torus. Now the definition (\ref{2.32}) implies
\begin{equation}
\label{2.36}
\psi (U_{1}\triangle U_{0}) + \psi (U_{2}\triangle U_{0}) =
\tau [\phi ](U_{1}) + \tau [\phi ](U_{2}) \; \hbox{mod} \, 2.
\end{equation}
It follows from the equalities (\ref{2.15}), (\ref{2.16}) and equality 
(\ref{2.20}) from Theorem 2.1 that the right hand side of the equality
(\ref{2.36}) equals $0$. Thus the values $\psi (U_{1}\triangle U_{0})$ 
and $\psi (U_{2}\triangle U_{0})$ coincide. The theorem is proved.

\noindent {\bf Theorem 2.4} {\it The symmetric difference}
$U\triangle U_{0}$ {\it of the dimer configurations} $U$, $U_{0}$ 
{\it on the graph} $cl(G(N,M))$ {\it generates all elements of the
homology group} $H_{1}(P(G)),{\bf Z}_{2}^{add})$.

\noindent {\it Proof.} Let the dimer configuration $U_{a}$ consist of the
edges $((3,j,1),(6,j,1))$, $((4,j,1),(5,j,1))$, $j = 1,...,N$;
$((1,j,1),(2,j + 1,1))$, $j = 1,...,N - 1$; $((1,N,1),(2,1,1))$;
$((1,j,k),(3,j,k))$, $((2,j,k),(4,j,k))$, $((5,j,k),(6,j,k))$,
$1 \leq j \leq N$, $2 \leq k \leq M$. Let us remind that the dimer
configuration $U_{0}$ corresponding to the cycle $\xi_{0}$ with the
coefficients in the group ${\bf Z}_{2}$ which takes value $1$ on all edges
of the graph $G(N,M)$ consists of the following edges of the graph
$cl(G(N,M))$: $((1,j,k),(3,j,k))$, $((2,j,k),(4,j,k))$,
$((5,j,k),(6,j,k))$, $1 \leq j \leq N$, $1 \leq k \leq M$. Thus the
symmetric difference $U_{a}\triangle U_{0}$ is the closed non - intersecting
itself non - homological to zero broken line which goes around the torus
horizontally. 

Let the dimer configuration $U_{b}$ consist of the edges: $((1,1,k),(6,1,k))$,
$((2,1,k),(5,1,k))$, $k = 1,...,M$; $((3,1,k),(4,1,k + 1))$,
$k = 1,...,M - 1$; $((3,1,M),(4,1,1))$; $((1,j,k),(3,j,k))$,
$((2,j,k),(4,j,k))$, $((5,j,k),(6,j,k))$, $2 \leq j \leq N$,
$1 \leq k \leq M$. Thus the symmetric difference
$U_{b}\triangle U_{0}$ is the closed non - intersecting itself 
non - homological to zero broken line which goes around the torus
vertically.

Let the dimer configuration $U_{a + b}$ consist of the edges: 
$((5,1,1),(6,1,1))$; $((3,j,1),(6,j,1))$, $((4,j,1),(5,j,1))$,
$2 \leq j \leq N$; $((1,1,k),(6,1,k))$, $((2,1,k),(5,1,k))$, 
$2 \leq k \leq M$; 

\noindent
$((1,j,1),(2,j + 1,1))$, $j = 1,...,N - 1$;
$((1,N,1),(2,1,1))$; $((3,1,k)(4,1,k + 1))$, $k = 1,...,M - 1$;
$((3,1,M),(4,1,1))$; $((1,j,k),(3,j,k))$, $((2,j,k)(4,j,k))$,
$((5,j,k),(6,j,k))$, $2 \leq j \leq N$, $2 \leq k \leq M$.
Thus the symmetric difference $U_{a + b}\triangle U_{0}$ is the closed
non - intersecting itself non - homological to zero broken line which
goes around the torus horizontally and vertically. 

At last $U_{0}\triangle U_{0} = \emptyset$. The equivalence classes of the
symmetric differences $U_{0}\triangle U_{0}$, $U_{a}\triangle U_{0}$,
$U_{b}\triangle U_{0}$, $U_{a + b}\triangle U_{0}$ as the closed curves on
the torus represent all elements of the homology group
$H_{1}(P(G),{\bf Z}_{2}^{add})$. The theorem is proved.

Let us define the cochain $a^{\ast} \in C^{1}(P(G),{\bf Z}_{2}^{add})$
in the following way: 

\noindent
$a^{\ast}(((1,N,k),(2,1,k))) = 1$, $1 \leq k \leq M$.
The cochain $a^{\ast}$ equals $0$ on all other edges of the graph 
$cl(G(N,M))$. By making use of the definition (\ref{2.28}) it is easy to
verify $\partial^{\ast} a^{\ast} = 0$ i. e. 
$a^{\ast} \in Z^{1}(P(G),{\bf Z}_{2}^{add})$. Due to definition
(\ref{2.29}) the linear function $\langle a^{\ast}, f\rangle$ is given
on the group $Z_{1}(P(G),{\bf Z}_{2}^{add})$. The equality (\ref{2.31})
and the equality $\partial^{\ast} a^{\ast} = 0$ imply that this function
is equal to zero on the group $B_{1}(P(G),{\bf Z}_{2}^{add})$.
Hence the linear function $\langle a^{\ast}, f\rangle$ defines the
function on the homology group $H_{1}(P(G),{\bf Z}_{2}^{add})$.
We denote by $U_{a}\triangle U_{0}$ the cochain which takes the value
$1$ on all edges of the broken line $U_{a}\triangle U_{0}$. It is equal to
zero on all other edges of the graph $cl(G(N,M))$. The cochains  
$U_{b}\triangle U_{0}$, $U_{a + b}\triangle U_{0}$ and $U_{0}\triangle U_{0}$
are defined analogously. The definition (\ref{2.29}) implies
\begin{eqnarray}
\label{2.37}
\langle a^{\ast},U_{a}\triangle U_{0}\rangle = 1 \; \hbox{mod} \, 2,
\langle a^{\ast},U_{b}\triangle U_{0}\rangle = 0 \; \hbox{mod} \, 2, 
\nonumber \\
\langle a^{\ast},U_{a + b}\triangle U_{0}\rangle = 1 \; \hbox{mod} \, 2,
\langle a^{\ast},U_{0}\triangle U_{0}\rangle = 0 \; \hbox{mod} \, 2.
\end{eqnarray}

We define the cochain $b^{\ast} \in C^{1}(P(G),{\bf Z}_{2}^{add})$ in the
following way: $b^{\ast}(((3,j,M),(4,j,1))) = 1$, $1 \leq j \leq N$.
The cochain $b^{\ast}$ equals $0$ on all other edges of the graph
$cl(G(N,M))$. By making use of the definition (\ref{2.28}) it is easy to
verify $\partial^{\ast} b^{\ast} = 0$ i. e.
$b^{\ast} \in Z^{1}(P(G),{\bf Z}_{2}^{add})$. Due to definition (\ref{2.29}) 
the linear function $\langle b^{\ast}, f\rangle$ is given on the group
$Z_{1}(P(G),{\bf Z}_{2}^{add})$. The equality (\ref{2.31}) and the equality
$\partial^{\ast} b^{\ast} = 0$ imply that this function is equal to zero on
the group $B_{1}(P(G),{\bf Z}_{2}^{add})$. Hence the linear function
$\langle b^{\ast}, f\rangle$ defines the function on the homology group
$H_{1}(P(G),{\bf Z}_{2}^{add})$. The definition (\ref{2.29}) implies
\begin{eqnarray}
\label{2.38}
\langle b^{\ast},U_{a}\triangle U_{0}\rangle = 0 \; \hbox{mod} \, 2,
\langle b^{\ast},U_{b}\triangle U_{0}\rangle = 1  \; \hbox{mod} \, 2, 
\nonumber \\
\langle b^{\ast},U_{a + b}\triangle U_{0}\rangle = 1 \; \hbox{mod} \, 2,
\langle b^{\ast},U_{0}\triangle U_{0}\rangle = 0 \; \hbox{mod} \, 2.
\end{eqnarray}

Let the orientation function $\phi$ satisfy the skew symmetry condition
(\ref{2.5}). Let us define three new orientation functions on the directed
edges
\begin{equation}
\label{2.39}
(\phi + a^{\ast})((p,q)) = \phi (p,q) + a^{\ast}(p,q) \; \hbox{mod} \, 2,
\end{equation}
\begin{equation}
\label{2.40}
(\phi + b^{\ast})((p,q)) = \phi (p,q) + b^{\ast}(p,q) \; \hbox{mod} \, 2,
\end{equation}
\begin{equation}
\label{2.41}
(\phi + a^{\ast} + b^{\ast})((p,q)) = \phi (p,q) + a^{\ast}(p,q) +
b^{\ast}(p,q) \; \hbox{mod} \, 2.
\end{equation}
Since $a^{\ast}(q,p) = a^{\ast}(p,q)$, $b^{\ast}(q,p) = b^{\ast}(p,q)$,
the orientation functions $\phi + a^{\ast}$, $\phi + b^{\ast}$, 
$\phi + a^{\ast} + b^{\ast}$ satisfy the skew symmetry condition (\ref{2.5}).

\noindent {\bf Theorem 2.5} {\it Let the orientation function} (\ref{2.18})
{\it be given on the graph} $cl(G(N,M))$. {\it The relation} (\ref{2.7}) 
{\it defines four antisymmetric} $6MN\times 6MN$ - {\it matrices:} 
$A(\phi)$, $A(\phi + a^{\ast})$, $A(\phi + b^{\ast})$, 
$A(\phi + a^{\ast} + b^{\ast})$. {\it Then the following relation for the
partition function} (\ref{2.2}) {\it holds}
\begin{eqnarray}
\label{2.42}
Z(N,M) = (- 2\cosh \beta E_{1}\cosh \beta E_{2})^{MN}
1/2 [- \hbox{Pf} \, A(\phi) + \hbox{Pf} \, A(\phi + a^{\ast}) + \nonumber \\
\hbox{Pf} \, A(\phi + b^{\ast}) + \hbox{Pf} \, A(\phi + a^{\ast} + b^{\ast})].
\end{eqnarray}
{\it Proof.} The relation (\ref{2.4}) may be rewritten as
\begin{equation}
\label{2.43}
Z(N,M) = (2\cosh \beta E_{1}\cosh \beta E_{2})^{MN}
\sum_{U} z(U)
\end{equation}
where the monomial $z(U)$  is given by the equality (\ref{2.12}) and the
summation in (\ref{2.43}) runs over all dimer configurations on the graph
$cl(G(N,M))$.

The equalities (\ref{2.10}), (\ref{2.18}) and (\ref{2.32}) define the
function $\psi (U\triangle U_{0})$ of the symmetric difference
$U\triangle U_{0}$ of the dimer configurations $U$, $U_{0}$ on the graph 
$cl(G(N,M))$. The dimer configuration $U_{0}$ corresponds with the cycle 
$\xi_{0}$ with the coefficients in the group ${\bf Z}_{2}$ which takes the
value $1$ on all edges of the graph $G(N,M)$. Due to Theorem 2.3 the function
$\psi (U\triangle U_{0})$ depends on the equivalence class of the cycle
$U\triangle U_{0}$ in the homology group $H_{1}(P(G),{\bf Z}_{2}^{add})$
only. Hence it is sufficient to calculate the function 
$\psi (U\triangle U_{0})$ on the following symmetric differences:
$U_{a}\triangle U_{0}$, $U_{b}\triangle U_{0}$, $U_{a + b}\triangle U_{0}$ 
and $U_{0}\triangle U_{0}$. It is easy to verify
\begin{equation}
\label{2.44}
\psi (U_{a}\triangle U_{0}) = \psi (U_{b}\triangle U_{0}) = 
\psi (U_{a + b}\triangle U_{0}) = 1 \; \hbox{mod} \, 2,
\psi (U_{0}\triangle U_{0}) = 0 \; \hbox{mod} \, 2.
\end{equation}  
By making use of the equalities (\ref{2.37}), (\ref{2.38}) and (\ref{2.44}) 
it is possible to verify the following relation 
\begin{equation}
\label{2.45}
(- 1)^{\psi (U\triangle U_{0})} = 1/2 (- 1 + 
(- 1)^{\langle a^{\ast},U\triangle U_{0}\rangle} +
(- 1)^{\langle b^{\ast},U\triangle U_{0}\rangle} +
(- 1)^{\langle a^{\ast},U\triangle U_{0}\rangle +
\langle b^{\ast},U\triangle U_{0}\rangle})
\end{equation}
for four symmetric differences: $U_{a}\triangle U_{0}$,
$U_{b}\triangle U_{0}$, $U_{a + b}\triangle U_{0}$, $U_{0}\triangle U_{0}$.
All terms of the equality (\ref{2.45}) depends on the equivalence class
of the cycle $U\triangle U_{0}$ in the homology group
$H_{1}(P(G),{\bf Z}_{2}^{add})$ only. Hence if the equality (\ref{2.45}) 
holds for the symmetric differences: $U_{a}\triangle U_{0}$,
$U_{b}\triangle U_{0}$, $U_{a + b}\triangle U_{0}$, $U_{0}\triangle U_{0}$,
it holds for the symmetric difference $U\triangle U_{0}$ where $U$ is an 
arbitrary dimer configuration on the graph $cl(G(N,M))$.

For any dimer configuration $U$ on the graph $cl(G(N,M))$ we introduce
the cochain $U \in C^{1}(P(G),{\bf Z}_{2}^{add})$ which equals $1$ on all
edges from the dimer configuration $U$ and which equals $0$ on all other
edges of the graph $cl(G(N,M))$. Due to definition the cochains $a^{\ast}$ 
and $b^{\ast}$ are equal to zero on all edges from the dimer configuration
$U_{0}$. Hence
\begin{equation}
\label{2.46}
\langle a^{\ast},U\triangle U_{0}\rangle =
\langle a^{\ast},U\rangle \; \hbox{mod} \, 2,
\langle b^{\ast},U\triangle U_{0}\rangle =
\langle b^{\ast},U\rangle \; \hbox{mod} \, 2.
\end{equation}
Let us multiply by $(- 1)^{\psi (U\triangle U_{0})}$ the equality
(\ref{2.45}). Then it follows from the equalities (\ref{2.32}),
(\ref{2.46}) and from Theorem 2.2 that
\begin{eqnarray}
\label{2.47}
1 = (- 1)^{MN}1/2[- (- 1)^{\tau [\phi ](U)} +
(- 1)^{\tau [\phi ](U) + \langle a^{\ast},U\rangle} + \nonumber \\
(- 1)^{\tau [\phi ](U) + \langle b^{\ast},U\rangle } +
(- 1)^{\tau [\phi ](U) + \langle a^{\ast},U\rangle +
\langle b^{\ast},U\rangle}].
\end{eqnarray}
Now substituting the relation (\ref{2.47}) into the equality 
(\ref{2.43}) and making use of the equalities (\ref{2.7}), (\ref{2.10}), 
(\ref{2.11}) we obtain the equality (\ref{2.42}). The theorem is proved.

For the calculation of the Pfaffians it is necessary to define the matrices 
$A(\phi)$, $A(\phi + a^{\ast})$, $A(\phi + b^{\ast})$,
$A(\phi + a^{\ast} + b^{\ast})$ explicitly. The relations
(\ref{2.5}) - (\ref{2.7}) and (\ref{2.18}) imply
\begin{eqnarray}
\label{2.48}
A(\phi )_{n(1,j_{1},k_{1}) n(2,j_{2},k_{2})} =
- A(\phi )_{n(2,j_{2},k_{2}) n(1,j_{1},k_{1})} = \nonumber \\
\tanh \beta E_{1}(\delta_{j_{1} + 1,j_{2}} + \delta_{j_{1}N} \delta_{j_{2}1})
\delta_{k_{1}k_{2}}
\end{eqnarray}
\begin{equation}
\label{2.49}
A(\phi )_{n(1,j_{1},k_{1}) n(3,j_{2},k_{2})} =
- A(\phi )_{n(3,j_{2},k_{2}) n(1,j_{1},k_{1})} = 
- \delta_{j_{1},j_{2}} \delta_{k_{1}k_{2}}
\end{equation}
\begin{equation}
\label{2.50}
A(\phi )_{n(1,j_{1},k_{1}) n(6,j_{2},k_{2})} =
- A(\phi )_{n(6,j_{2},k_{2}) n(1,j_{1},k_{1})} = 
\delta_{j_{1},j_{2}} \delta_{k_{1}k_{2}}
\end{equation}
\begin{equation}
\label{2.51}
A(\phi )_{n(2,j_{1},k_{1}) n(4,j_{2},k_{2})} =
- A(\phi )_{n(4,j_{2},k_{2}) n(2,j_{1},k_{1})} = 
- \delta_{j_{1},j_{2}} \delta_{k_{1}k_{2}}
\end{equation}
\begin{equation}
\label{2.52}
A(\phi )_{n(2,j_{1},k_{1}) n(5,j_{2},k_{2})} =
- A(\phi )_{n(5,j_{2},k_{2}) n(2,j_{1},k_{1})} = 
\delta_{j_{1},j_{2}} \delta_{k_{1}k_{2}}
\end{equation}
\begin{eqnarray}
\label{2.53}
A(\phi )_{n(3,j_{1},k_{1}) n(4,j_{2},k_{2})} =
- A(\phi )_{n(4,j_{2},k_{2}) n(3,j_{1},k_{1})} = \nonumber \\
\tanh \beta E_{2} \delta_{j_{1}j_{2}}
(\delta_{k_{1} + 1,k_{2}} + \delta_{k_{1}M} \delta_{k_{2}1})
\end{eqnarray}
\begin{equation}
\label{2.54}
A(\phi )_{n(3,j_{1},k_{1}) n(6,j_{2},k_{2})} =
- A(\phi )_{n(6,j_{2},k_{2}) n(3,j_{1},k_{1})} = 
- \delta_{j_{1},j_{2}} \delta_{k_{1}k_{2}}
\end{equation}
\begin{equation}
\label{2.55}
A(\phi )_{n(4,j_{1},k_{1}) n(5,j_{2},k_{2})} =
- A(\phi )_{n(5,j_{2},k_{2}) n(4,j_{1},k_{1})} = 
- \delta_{j_{1},j_{2}} \delta_{k_{1}k_{2}}
\end{equation}
\begin{equation}
\label{2.56}
A(\phi )_{n(5,j_{1},k_{1}) n(6,j_{2},k_{2})} =
- A(\phi )_{n(6,j_{2},k_{2}) n(5,j_{1},k_{1})} = 
\delta_{j_{1},j_{2}} \delta_{k_{1}k_{2}}
\end{equation}
All other elements of the matrix $A(\phi)$ are equal to zero.

In view of the definition the cochain $a^{\ast}$ equals $1$ on the edges
$((1,N,k),(2,1,k))$, $1 \leq k \leq M$ and it is equal to zero on all
other edges. The equalities (\ref{2.5}) - (\ref{2.7}) and (\ref{2.18})
imply
\begin{eqnarray}
\label{2.57}
A(\phi + a^{\ast})_{n(1,j_{1},k_{1}) n(2,j_{2},k_{2})} =
- A(\phi + a^{\ast})_{n(2,j_{2},k_{2}) n(1,j_{1},k_{1})} = \nonumber \\
\tanh \beta E_{1}(\delta_{j_{1} + 1,j_{2}} - \delta_{j_{1}N} \delta_{j_{2}1})
\delta_{k_{1}k_{2}}
\end{eqnarray}
All the other elements of the matrices $A(\phi)$ and $A(\phi + a^{\ast})$
coincide. These non - zero elements are given by the relations
(\ref{2.49}) - (\ref{2.56}).

Due to definition the cochain $b^{\ast}$ equals $1$ on the edges
$((3,j,M),(4,j,1))$, $1 \leq j \leq N$, and it is equal to zero on all other 
edges. The equalities (\ref{2.5}) - (\ref{2.7}) and (\ref{2.18}) imply
\begin{eqnarray}
\label{2.58}
A(\phi + b^{\ast})_{n(3,j_{1},k_{1}) n(4,j_{2},k_{2})} =
- A(\phi + b^{\ast})_{n(4,j_{2},k_{2}) n(3,j_{1},k_{1})} = \nonumber \\
\tanh \beta E_{2} \delta_{j_{1}j_{2}}
(\delta_{k_{1} + 1,k_{2}} - \delta_{k_{1}M} \delta_{k_{2}1})
\end{eqnarray}
All the other elements of the matrices $A(\phi)$ and $A(\phi + b^{\ast})$
coincide. These non - zero elements are given by the relations
(\ref{2.48}) - (\ref{2.52}), (\ref{2.54}) - (\ref{2.56}).

In view of the relations (\ref{2.5}) - (\ref{2.7}) and (\ref{2.18}) 
and the definitions of the cochains $a^{\ast}$, $b^{\ast}$ we have
\begin{eqnarray}
\label{2.59}
A(\phi + a^{\ast} + b^{\ast})_{n(1,j_{1},k_{1}) n(2,j_{2},k_{2})} =
- A(\phi + a^{\ast} + b^{\ast})_{n(2,j_{2},k_{2}) n(1,j_{1},k_{1})} = 
\nonumber \\
\tanh \beta E_{1}(\delta_{j_{1} + 1,j_{2}} - \delta_{j_{1}N} \delta_{j_{2}1})
\delta_{k_{1}k_{2}}
\end{eqnarray}
\begin{eqnarray}
\label{2.60}
A(\phi + a^{\ast} + b^{\ast})_{n(3,j_{1},k_{1}) n(4,j_{2},k_{2})} =
- A(\phi + a^{\ast} + b^{\ast})_{n(4,j_{2},k_{2}) n(3,j_{1},k_{1})} = 
\nonumber \\
\tanh \beta E_{2} \delta_{j_{1}j_{2}}
(\delta_{k_{1} + 1,k_{2}} - \delta_{k_{1}M} \delta_{k_{2}1})
\end{eqnarray}
All the other elements of the matrices $A(\phi)$ and 
$A(\phi + a^{\ast} + b^{\ast})$ coincide. These non - zero elements are
given by the relations (\ref{2.49}) - (\ref{2.52}), (\ref{2.54}) - 
(\ref{2.56}).

In the next section we calculate Pfaffians of the matrices (\ref{2.48}) - 
(\ref{2.60}).

\section{Pfaffians}
\setcounter{equation}{0}

For the natural numbers $- N < j \leq N$ the following relation is valid
\begin{equation}
\label{3.1}
N^{- 1}\sum_{j^{\prime} = 1}^{N} \exp \{ i2\pi N^{- 1}jj^{\prime}\} =
\delta_{j0} + \delta_{jN}.
\end{equation}
The relations (\ref{2.48}) - (\ref{2.56}) and (\ref{3.1}) imply
\begin{eqnarray}
\label{3.2}
A(\phi )_{n(i_{1},j_{1},k_{1}) n(i_{2},j_{2},k_{2})} =
\sum_{i_{1}^{\prime},i_{2}^{\prime} = 1}^{6}
\sum_{j_{1}^{\prime},j_{2}^{\prime} = 1}^{N}
\sum_{k_{1}^{\prime},k_{2}^{\prime} = 1}^{M} \nonumber \\
B(\phi )_{n(i_{1}^{\prime},j_{1}^{\prime},k_{1}^{\prime}) n(i_{1},j_{1},k_{1})}
\tilde{A} (\phi )_{n(i_{1}^{\prime},j_{1}^{\prime},k_{1}^{\prime}) 
n(i_{2}^{\prime},j_{2}^{\prime},k_{2}^{\prime})}
B(\phi )_{n(i_{2}^{\prime},j_{2}^{\prime},k_{2}^{\prime}) n(i_{2},j_{2},k_{2})}
\end{eqnarray}
where
\begin{equation}
\label{3.3}
B(\phi )_{n(p^{\prime},j^{\prime},k^{\prime}) n(p,j,k)} =
\delta_{pp^{\prime}}
\exp \{ i2\pi (N^{- 1}jj^{\prime} + M^{- 1}kk^{\prime})\}
\end{equation}
\begin{eqnarray}
\label{3.4}
\tilde{A} (\phi )_{n(1,j_{1},k_{1}) n(2,j_{2},k_{2})} =
- \tilde{A} (\phi )_{n(2,j_{2},k_{2}) n(1,j_{1},k_{1})} = \nonumber \\
\tanh \beta E_{1}\exp\{ i2\pi N^{- 1}j_{1}\}
\delta (j_{1},j_{2},k_{1},k_{2})
\end{eqnarray}
\begin{eqnarray}
\label{3.5}
\tilde{A} (\phi )_{n(3,j_{1},k_{1}) n(4,j_{2},k_{2})} =
- \tilde{A} (\phi )_{n(4,j_{2},k_{2}) n(3,j_{1},k_{1})} = \nonumber \\
\tanh \beta E_{2}\exp\{ i2\pi M^{- 1}k_{1}\}
\delta (j_{1},j_{2},k_{1},k_{2})
\end{eqnarray}
\begin{eqnarray}
\label{3.6}
\tilde{A} (\phi )_{n(1,j_{1},k_{1}) n(3,j_{2},k_{2})} =
- \tilde{A} (\phi )_{n(3,j_{2},k_{2}) n(1,j_{1},k_{1})} = \nonumber \\
\tilde{A} (\phi )_{n(1,j_{1},k_{1}) n(6,j_{2},k_{2})} =
- \tilde{A} (\phi )_{n(6,j_{2},k_{2}) n(1,j_{1},k_{1})} = \nonumber \\
- \tilde{A} (\phi )_{n(2,j_{1},k_{1}) n(4,j_{2},k_{2})} =
\tilde{A} (\phi )_{n(4,j_{2},k_{2}) n(2,j_{1},k_{1})} = \nonumber \\
\tilde{A} (\phi )_{n(2,j_{1},k_{1}) n(5,j_{2},k_{2})} =
- \tilde{A} (\phi )_{n(5,j_{2},k_{2}) n(2,j_{1},k_{1})} = \nonumber \\
- \tilde{A} (\phi )_{n(3,j_{1},k_{1}) n(6,j_{2},k_{2})} =
\tilde{A} (\phi )_{n(6,j_{2},k_{2}) n(3,j_{1},k_{1})} = \nonumber \\
- \tilde{A} (\phi )_{n(4,j_{1},k_{1}) n(5,j_{2},k_{2})} =
\tilde{A} (\phi )_{n(5,j_{2},k_{2}) n(4,j_{1},k_{1})} = \nonumber \\
\tilde{A} (\phi )_{n(5,j_{1},k_{1}) n(6,j_{2},k_{2})} =
- \tilde{A} (\phi )_{n(6,j_{2},k_{2}) n(5,j_{1},k_{1})} = \nonumber \\
\delta (j_{1},j_{2},k_{1},k_{2})
\end{eqnarray}
\begin{equation}
\label{3.7}
\delta (j_{1},j_{2},k_{1},k_{2}) =
(MN)^{- 1}(\delta_{N - j_{1},j_{2}} + \delta_{j_{1}N} \delta_{j_{2}N})
(\delta_{M - k_{1},k_{2}} + \delta_{k_{1}M} \delta_{k_{2}M})
\end{equation}
All the other elements of the matrix $\tilde{A} (\phi)$ are equal to zero.

Due to Proposition 1 from (\cite{12}, chap. IX, sect. 5)
\begin{equation}
\label{3.8}
\hbox{Pf} \, (B^{T}\tilde{A}(\phi)B) = \det B \hbox{Pf} \, \tilde{A}(\phi).
\end{equation}

In view of the definition (\ref{3.3}) the matrix 
$B = I_{6}\otimes C_{N}\otimes C_{M}$, where $I_{6}$ is the identity 
$6\times 6$ - matrix and $N\times N$ - matrix
$(C_{N})_{kj} = \exp \{ i 2\pi N^{- 1}kj\}$. Therefore
\begin{equation}
\label{3.9}
\det B = (\det C_{N})^{6M}(\det C_{M})^{6N}. 
\end{equation}
For the determinant of the matrix 
$(\bar{C}_{N})_{jk} = \exp \{ i2\pi N^{- 1}(N - j)k\}$ we have
\begin{equation}
\label{3.10}
\det \bar{C}_{N} = (- 1)^{[\frac{N - 1}{2}]}\det C_{N}
\end{equation}
where $[r]$ denotes the integral part of the real number $r$. The relation 
(\ref{3.1}) implies
\begin{equation}
\label{3.11}
\bar{C}_{N}C_{N} = NI_{N}
\end{equation}
where $I_{N}$ is the identity $N\times N$ - matrix. It follows from the
relations (\ref{3.10}), (\ref{3.11}) 
\begin{equation}
\label{3.12}
(\det C_{N})^{2} = (- 1)^{[\frac{N - 1}{2}]}N^{N}.
\end{equation}
The substitution of the equality (\ref{3.12}) into the equality (\ref{3.9})
yeilds
\begin{equation}
\label{3.13}
\det B = (- 1)^{M[\frac{N - 1}{2}] + [\frac{M - 1}{2}]N}(MN)^{3MN}.
\end{equation}

Instead of the definition (\ref{2.8}) we make use of the definition from
(\cite{12}, chap. IX, sect. 5). Let $e_{k}$, $k = 1,...,6MN$, be the basis
of the space where the matrix $\tilde{A}(\phi)$ acts. Then
\begin{equation}
\label{3.14}
\wedge^{3MN} \left( \sum_{j,k = 1}^{6MN} \tilde{A}(\phi)_{jk} 
e_{j}\wedge e_{k} \right) =
2^{3MN}(3MN)!\hbox{Pf} \, \tilde{A}(\phi) e_{1}\wedge \cdots \wedge e_{6MN}.
\end{equation}

We introduce $6\times 6$ - matrix
\begin{eqnarray}
\label{3.15}
(D(j,k))_{i_{1}i_{2}} = 
MN \tilde{A}(\phi)_{n(i_{1},j,k),n(i_{2},N - j,M - k)} \nonumber \\
(D(N,k))_{i_{1}i_{2}} =
MN \tilde{A}(\phi)_{n(i_{1},N,k),n(i_{2},N,M - k)} \nonumber \\
(D(j,M))_{i_{1}i_{2}} =
MN \tilde{A}(\phi)_{n(i_{1},j,M),n(i_{2},N - j,M)} \nonumber \\
(D(N,M))_{i_{1}i_{2}} =
MN \tilde{A}(\phi)_{n(i_{1},N,M),n(i_{2},N,M)}
\end{eqnarray}
where $1 \leq i_{1} \leq 6$, $1 \leq i_{2} \leq 6$, $1 \leq j < N$,
$1 \leq k < M$. In view of the definitions (\ref{3.4}) - (\ref{3.7})
the matrix (\ref{3.15}) is anti - Hermitian:
$(D(j,k))_{i_{2}i_{1}} = - \overline{(D(j,k))_{i_{1}i_{2}}}$. For the natural
numbers $j = \frac{N}{2}, N$, $k = \frac{M}{2}, M$ the matrix $D(j,k)$ 
is real and therefore antisymmetric. Hence its Pfaffian is defined for 
these natural numbers $j = \frac{N}{2}, N$, $k = \frac{M}{2}, M$
\begin{eqnarray}
\label{3.16}
\hbox{Pf} \, D(j,k) = \nonumber \\
z_{1}z_{2}\exp \{i2\pi (N^{- 1}j + M^{- 1}k)\} +
z_{1}\exp \{i2\pi N^{- 1}j\} + z_{2}\exp \{i2\pi M^{- 1}k\} - 1
\end{eqnarray}
where $z_{i} = \tanh \beta E_{i}$, $i = 1,2$. The determinant of the
matrix $D(j,k)$ is defined for any numbers $1 \leq j \leq N$, 
$1 \leq k \leq M$
\begin{eqnarray}
\label{3.17}
\det D(j,k) = \nonumber \\
(1 + z_{1}^{2})(1 + z_{2}^{2}) - 2z_{1}(1 - z_{2}^{2})\cos (2\pi N^{- 1}j)
- 2z_{2}(1 - z_{1}^{2})\cos (2\pi M^{- 1}k).
\end{eqnarray}
The equality (\ref{1.4}) coincides with the relation
\begin{equation}
\label{3.18}
\det D(N,M) = (\hbox{Pf} \, D(N,M))^{2}.
\end{equation}

By making use of the definitions (\ref{3.4}) - (\ref{3.7}) and the Pfaffian
definition (\ref{3.14}) it is possible to show that the Pfaffian of the
$6MN\times 6MN$ - matrix $\tilde{A}(\phi)$ is proportional to the product
of all Pfaffians of $6\times 6$ - matrices $D(j,k)$ for the natural 
numbers $j = \frac{N}{2}, N$, $k = \frac{M}{2}, M$ and the product of all 
determinants of the matrices $D(j,k)$, $1 \leq j \leq N$, $1 \leq k \leq M$, 
except $j = \frac{N}{2}, N$, $k = \frac{M}{2}, M$ simultaneously.

Let us introduce the products
\begin{eqnarray}
\label{3.19}
P_{11}(l_{1},l_{2};m_{1},m_{2}) = \prod_{j = l_{1}}^{l_{2}}
\prod_{k = m_{1}}^{m_{2}} [(1 + z_{1}^{2})(1 + z_{2}^{2}) - \nonumber \\
- 2z_{1}(1 - z_{2}^{2})\cos (2\pi N^{- 1}j)
- 2z_{2}(1 - z_{1}^{2})\cos (2\pi M^{- 1}k)]
\end{eqnarray}
\begin{eqnarray}
\label{3.20}
P_{21}(l_{1},l_{2};m_{1},m_{2}) = \prod_{j = l_{1}}^{l_{2}}
\prod_{k = m_{1}}^{m_{2}} [(1 + z_{1}^{2})(1 + z_{2}^{2}) - \nonumber \\
- 2z_{1}(1 - z_{2}^{2})\cos (\pi N^{- 1}(2j - 1))
- 2z_{2}(1 - z_{1}^{2})\cos (2\pi M^{- 1}k)]
\end{eqnarray}
\begin{eqnarray}
\label{3.21}
P_{12}(l_{1},l_{2};m_{1},m_{2}) = \prod_{j = l_{1}}^{l_{2}}
\prod_{k = m_{1}}^{m_{2}} [(1 + z_{1}^{2})(1 + z_{2}^{2}) - \nonumber \\
- 2z_{1}(1 - z_{2}^{2})\cos (2\pi N^{- 1}j)
- 2z_{2}(1 - z_{1}^{2})\cos (\pi M^{- 1}(2k - 1))]
\end{eqnarray}
\begin{eqnarray}
\label{3.22}
P_{22}(l_{1},l_{2};m_{1},m_{2}) = \prod_{j = l_{1}}^{l_{2}}
\prod_{k = m_{1}}^{m_{2}} [(1 + z_{1}^{2})(1 + z_{2}^{2}) - \nonumber \\
- 2z_{1}(1 - z_{2}^{2})\cos (\pi N^{- 1}(2j - 1))
- 2z_{2}(1 - z_{1}^{2})\cos (\pi M^{- 1}(2k - 1))].
\end{eqnarray}

Calculating the Pfaffian $\hbox{Pf} \, \tilde{A}(\phi)$ and making use
of the relations (\ref{3.2}), (\ref{3.8}) and (\ref{3.13}) we have:

\noindent if the numbers $N,M$ are odd, then
\begin{equation}
\label{3.23}
\hbox{Pf} \, A(\phi) = F_{11}(z_{1},z_{2})(z_{1}z_{2} + z_{1} + z_{2} - 1),
\end{equation}
where $z_{i} = \tanh \beta E_{i}$, $i = 1,2$ and
\begin{equation}
\label{3.24}
F_{11}(z_{1},z_{2}) = 
P_{11}(0,\frac{N - 1}{2}; 1,\frac{M - 1}{2})
P_{11}(1,\frac{N - 1}{2}; 0,\frac{M - 1}{2}); 
\end{equation}
if the number $N$ is odd and the number $M$ is even, then
\begin{equation}
\label{3.25}
\hbox{Pf} \, A(\phi) = F_{12}(z_{1},z_{2})(- z_{1}z_{2} + z_{1} - z_{2} - 1)
(z_{1}z_{2} + z_{1} + z_{2} - 1),
\end{equation}
\begin{equation}
\label{3.26}
F_{12}(z_{1},z_{2}) = 
P_{11}(0,\frac{N - 1}{2}; 1,\frac{M}{2} - 1)
P_{11}(1,\frac{N - 1}{2}; 0,\frac{M}{2});
\end{equation} 
if the number $N$ is even and the number $M$ is odd, then 
\begin{equation}
\label{3.27}
\hbox{Pf} \, A(\phi) = F_{13}(z_{1},z_{2})(- z_{1}z_{2} - z_{1} + z_{2} - 1)
(z_{1}z_{2} + z_{1} + z_{2} - 1),
\end{equation}
\begin{equation}
\label{3.28}
F_{13}(z_{1},z_{2}) = 
P_{11}(0,\frac{N}{2}; 1,\frac{M - 1}{2})
P_{11}(1,\frac{N}{2} - 1;0,\frac{M - 1}{2});
\end{equation}
if the numbers $N,M$ are even, then
\begin{eqnarray}
\label{3.29}
\hbox{Pf} \, A(\phi) = F_{14}(z_{1},z_{2})(z_{1}z_{2} - z_{1} - z_{2} - 1)
(- z_{1}z_{2} - z_{1} + z_{2} - 1) \times \nonumber \\
(- z_{1}z_{2} + z_{1} - z_{2} - 1)(z_{1}z_{2} + z_{1} + z_{2} - 1),
\end{eqnarray}
\begin{equation}
\label{3.30}
F_{14}(z_{1},z_{2}) = 
P_{11}(0,\frac{N}{2}; 1,\frac{M}{2} - 1)
P_{11}(1,\frac{N}{2} - 1;0,\frac{M}{2}).
\end{equation}

Analogously the definitions (\ref{2.49}) - (\ref{2.57}) imply:

\noindent if the numbers $N,M$ are odd, then
\begin{equation}
\label{3.31}
\hbox{Pf} \, A(\phi + a^{\ast}) = F_{21}(z_{1},z_{2})
(- z_{1}z_{2} - z_{1} + z_{2} - 1),
\end{equation}
\begin{equation}
\label{3.32}
F_{21}(z_{1},z_{2}) = 
P_{21}(1,\frac{N + 1}{2}; 1,\frac{M - 1}{2})
P_{21}(1,\frac{N - 1}{2}; 0,\frac{M - 1}{2});
\end{equation}
if the number $N$ is odd and the number $M$ is even, then
\begin{equation}
\label{3.33}
\hbox{Pf} \, A(\phi + a^{\ast}) = 
F_{22}(z_{1},z_{2})(z_{1}z_{2} - z_{1} - z_{2} - 1)
(- z_{1}z_{2} - z_{1} + z_{2} - 1),
\end{equation}
\begin{equation}
\label{3.34}
F_{22}(z_{1},z_{2}) = 
P_{21}(1,\frac{N + 1}{2}; 1,\frac{M}{2} - 1)
P_{21}(1,\frac{N - 1}{2}; 0,\frac{M}{2});
\end{equation}
if the number $N$ is even and the number $M$ is odd, then
\begin{eqnarray}
\label{3.35}
\hbox{Pf} \, A(\phi + a^{\ast}) = F_{23}(z_{1},z_{2}) = \nonumber \\
P_{21}(1,\frac{N}{2}; 1,\frac{M - 1}{2})
P_{21}(1,\frac{N}{2}; 0,\frac{M - 1}{2});
\end{eqnarray}
if the numbers $N,M$ are even, then
\begin{eqnarray}
\label{3.36}
\hbox{Pf} \, A(\phi + a^{\ast}) = F_{24}(z_{1},z_{2}) = \nonumber \\
P_{21}(1,\frac{N}{2}; 1,\frac{M}{2})
P_{21}(1,\frac{N}{2}; 0,\frac{M}{2} - 1).
\end{eqnarray}

Similarly the definitions (\ref{2.48}) - (\ref{2.52}),
(\ref{2.54}) - (\ref{2.56}) and (\ref{2.58}) imply:

\noindent if the numbers $N,M$ are odd, then
\begin{equation}
\label{3.37}
\hbox{Pf} \, A(\phi + b^{\ast}) = 
F_{31}(z_{1},z_{2})(- z_{1}z_{2} + z_{1} - z_{2} - 1),
\end{equation}
\begin{equation}
\label{3.38}
F_{31}(z_{1},z_{2}) = 
P_{12}(0,\frac{N - 1}{2}; 1,\frac{M - 1}{2})
P_{12}(1,\frac{N - 1}{2}; 1,\frac{M + 1}{2});
\end{equation} 
if the number $N$ is odd and the number $M$ is even, then
\begin{eqnarray}
\label{3.39}
\hbox{Pf} \, A(\phi + b^{\ast}) = F_{32}(z_{1},z_{2}) = \nonumber \\
P_{12}(0,\frac{N - 1}{2}; 1,\frac{M}{2})
P_{12}(1,\frac{N - 1}{2}; 1,\frac{M}{2});
\end{eqnarray}
if the number $N$ is even and the number $M$ is odd, then 
\begin{equation}
\label{3.40}
\hbox{Pf} \, A(\phi + b^{\ast}) = 
F_{33}(z_{1},z_{2})(z_{1}z_{2} - z_{1} - z_{2} - 1)
(- z_{1}z_{2} + z_{1} - z_{2} - 1),
\end{equation}
\begin{equation}
\label{3.41}
F_{33}(z_{1},z_{2}) = 
P_{12}(0,\frac{N}{2}; 1,\frac{M - 1}{2})
P_{12}(1,\frac{N}{2} - 1;1,\frac{M + 1}{2});
\end{equation}
if the numbers $N,M$ are even, then
\begin{eqnarray}
\label{3.42}
\hbox{Pf} \, A(\phi + b^{\ast}) = F_{34}(z_{1},z_{2}) = \nonumber \\
P_{12}(0,\frac{N}{2} - 1;1,\frac{M}{2})
P_{12}(1,\frac{N}{2}; 1,\frac{M}{2}).
\end{eqnarray}

Similarly the definitions (\ref{2.49}) - (\ref{2.52}),
(\ref{2.54}) - (\ref{2.56}), (\ref{2.59}), (\ref{2.60}) imply:

\noindent if the numbers $N,M$ are odd, then
\begin{equation}
\label{3.43}
\hbox{Pf} \, A(\phi + a^{\ast} + b^{\ast}) = F_{41}(z_{1},z_{2})
(z_{1}z_{2} - z_{1} - z_{2} - 1),
\end{equation}
\begin{equation}
\label{3.44}
F_{41}(z_{1},z_{2}) = 
P_{22}(1,\frac{N + 1}{2}; 1,\frac{M - 1}{2})
P_{22}(1,\frac{N - 1}{2}; 1,\frac{M + 1}{2});
\end{equation}
if the number $N$ is odd and the number $M$ is even, then
\begin{eqnarray}
\label{3.45}
\hbox{Pf} \, A(\phi + a^{\ast} + b^{\ast}) = F_{42}(z_{1},z_{2}) = \nonumber \\
P_{22}(1,\frac{N + 1}{2}; 1,\frac{M}{2})
P_{22}(1,\frac{N - 1}{2}; 1,\frac{M}{2});
\end{eqnarray}
if the number $N$ is even and the number $M$ is odd, then
\begin{eqnarray}
\label{3.46}
\hbox{Pf} \, A(\phi + a^{\ast} + b^{\ast}) = F_{43}(z_{1},z_{2}) = \nonumber \\
P_{22}(1,\frac{N}{2}; 1,\frac{M - 1}{2})
P_{22}(1,\frac{N}{2}; 1,\frac{M + 1}{2});
\end{eqnarray}
if the numbers $N,M$ are even, then
\begin{eqnarray}
\label{3.47}
\hbox{Pf} \, A(\phi + a^{\ast} + b^{\ast}) = F_{44}(z_{1},z_{2}) = \nonumber \\
(P_{22}(1,\frac{N}{2}; 1,\frac{M}{2} ))^{2}.
\end{eqnarray}

The formulae (\ref{3.23}) - (\ref{3.47}) are in accordance with the
formulae \cite{9} for $\det \bar{A}_{i}$. The formulae (\ref{1.4}),
(\ref{3.23}) - (\ref{3.47}) imply
\begin{eqnarray}
\label{3.48}
\det A(\phi) = P_{11}(1,N;1,M) \nonumber \\
\det A(\phi + a^{\ast}) = P_{21}(1,N;1,M) \nonumber \\
\det A(\phi + b^{\ast}) = P_{12}(1,N;1,M) \nonumber \\
\det A(\phi + a^{\ast} + b^{\ast}) = P_{22}(1,N;1,M).
\end{eqnarray}
The expressions (\ref{3.48}) coicide with the expressions  for
$\det \bar{A}_{4}$, $i = 1,...,4$, given in the book \cite{9}.

\section{Free Energy}
\setcounter{equation}{0}

The following lemma is the staightforward consequence of the relation  
(\ref{1.4}).

\noindent {\bf Lemma 4.1} {\it For the variables} 
$z_{i} = \tanh \beta E_{i}$, $\beta > 0$, $E_{i} \neq 0$, $i = 1,2$ 
{\it the polynomial} 
\begin{equation}
\label{4.1}
(1 + z_{1}^{2})(1 + z_{2}^{2}) - 2z_{1}(1 - z_{2}^{2})\cos \phi_{1}
- 2z_{2}(1 - z_{1}^{2})\cos \phi_{2} > 0
\end{equation}
{\it if at least one of the followng conditions} 
\begin{equation}
\label{4.2}
\cos \phi_{i} = \hbox{sgn} \, z_{i}, i = 1,2,
\end{equation}
{\it is not valid.}

\noindent {\it Proof.} Since $0 < |z_{i}| < 1$, $i = 1,2$, the following
inequality holds
\begin{eqnarray}
\label{4.3}
(1 + z_{1}^{2})(1 + z_{2}^{2}) - 2z_{1}(1 - z_{2}^{2})\cos \phi_{1}
- 2z_{2}(1 - z_{1}^{2})\cos \phi_{2} \geq \nonumber \\
(1 + z_{1}^{2})(1 + z_{2}^{2}) - 2|z_{1}|(1 - z_{2}^{2})
- 2|z_{2}|(1 - z_{1}^{2}) 
\end{eqnarray}
Due to the equality (\ref{1.4}) the right hand side of the inequality
(\ref{4.3}) is equal to
\begin{equation}
\label{4.4}
(|z_{1}z_{2}| + |z_{1}| + |z_{2}| - 1)^{2}.
\end{equation}
The inequality (\ref{4.3}) may be an equality if and only if both
conditions (\ref{4.2}) are valid. The lemma is proved.

\noindent {\bf Lemma 4.2} {\it The range of values for the variables}
$z_{i} = \tanh \beta E_{i}$, $\beta > 0$, $E_{i} \neq 0$, $i = 1,2$ 
{\it is divided into eight two dimensional domains:} 
\begin{eqnarray}
\label{4.5}
(1 + z_{1})(1 + z_{2}) > 2,(1 - z_{1})(1 + z_{2}) < 2, \nonumber \\
(1 + z_{1})(1 - z_{2}) < 2, (1 - z_{1})(1 - z_{2}) < 2, \nonumber \\
z_{1} > 0, z_{2} > 0;
\end{eqnarray}
\begin{eqnarray}
\label{4.6}
(1 + z_{1})(1 + z_{2}) < 2,(1 - z_{1})(1 + z_{2}) < 2, \nonumber \\
(1 + z_{1})(1 - z_{2}) < 2, (1 - z_{1})(1 - z_{2}) < 2, \nonumber \\
z_{1} > 0, z_{2} > 0;
\end{eqnarray}
\begin{eqnarray}
\label{4.7}
(1 + z_{1})(1 + z_{2}) < 2,(1 - z_{1})(1 + z_{2}) > 2, \nonumber \\
(1 + z_{1})(1 - z_{2}) < 2, (1 - z_{1})(1 - z_{2}) < 2, \nonumber \\
z_{1} < 0, z_{2} > 0;
\end{eqnarray}
\begin{eqnarray}
\label{4.8}
(1 + z_{1})(1 + z_{2}) < 2,(1 - z_{1})(1 + z_{2}) < 2, \nonumber \\
(1 + z_{1})(1 - z_{2}) < 2, (1 - z_{1})(1 - z_{2}) < 2, \nonumber \\
z_{1} < 0, z_{2} > 0;
\end{eqnarray}
\begin{eqnarray}
\label{4.9}
(1 + z_{1})(1 + z_{2}) < 2,(1 - z_{1})(1 + z_{2}) < 2, \nonumber \\
(1 + z_{1})(1 - z_{2}) > 2, (1 - z_{1})(1 - z_{2}) < 2, \nonumber \\
z_{1} > 0, z_{2} < 0;
\end{eqnarray}
\begin{eqnarray}
\label{4.10}
(1 + z_{1})(1 + z_{2}) < 2,(1 - z_{1})(1 + z_{2}) < 2, \nonumber \\
(1 + z_{1})(1 - z_{2}) < 2, (1 - z_{1})(1 - z_{2}) < 2, \nonumber \\
z_{1} > 0, z_{2} < 0;
\end{eqnarray}
\begin{eqnarray}
\label{4.11}
(1 + z_{1})(1 + z_{2}) < 2,(1 - z_{1})(1 + z_{2}) < 2, \nonumber \\
(1 + z_{1})(1 - z_{2}) < 2, (1 - z_{1})(1 - z_{2}) > 2, \nonumber \\
z_{1} < 0, z_{2} < 0;
\end{eqnarray}
\begin{eqnarray}
\label{4.12}
(1 + z_{1})(1 + z_{2}) < 2,(1 - z_{1})(1 + z_{2}) < 2, \nonumber \\
(1 + z_{1})(1 - z_{2}) < 2, (1 - z_{1})(1 - z_{2}) < 2, \nonumber \\
z_{1} < 0, z_{2} < 0;
\end{eqnarray}
{\it and into four one dimensional domains:}
\begin{eqnarray}
\label{4.13}
(1 + z_{1})(1 + z_{2}) = 2,(1 - z_{1})(1 + z_{2}) < 2, \nonumber \\
(1 + z_{1})(1 - z_{2}) < 2, (1 - z_{1})(1 - z_{2}) < 2, \nonumber \\
z_{1} > 0, z_{2} > 0;
\end{eqnarray}
\begin{eqnarray}
\label{4.14}
(1 + z_{1})(1 + z_{2}) < 2,(1 - z_{1})(1 + z_{2}) = 2, \nonumber \\
(1 + z_{1})(1 - z_{2}) < 2, (1 - z_{1})(1 - z_{2}) < 2, \nonumber \\
z_{1} < 0, z_{2} > 0;
\end{eqnarray}
\begin{eqnarray}
\label{4.15}
(1 + z_{1})(1 + z_{2}) < 2,(1 - z_{1})(1 + z_{2}) < 2, \nonumber \\
(1 + z_{1})(1 - z_{2}) = 2, (1 - z_{1})(1 - z_{2}) < 2, \nonumber \\
z_{1} > 0, z_{2} < 0;
\end{eqnarray}
\begin{eqnarray}
\label{4.16}
(1 + z_{1})(1 + z_{2}) < 2,(1 - z_{1})(1 + z_{2}) < 2, \nonumber \\
(1 + z_{1})(1 - z_{2}) < 2, (1 - z_{1})(1 - z_{2}) = 2, \nonumber \\
z_{1} < 0, z_{2} < 0.
\end{eqnarray} 
{\it Proof.} Let two sets of numbers $\epsilon_{i} = \pm 1$, 
$\sigma_{i} = \pm 1$, $i = 1,2$ be given. Let us prove that if these two
sets do not coinside, then two inequalities  
\begin{eqnarray}
\label{4.17}
(1 + \epsilon_{1} z_{1})(1 + \epsilon_{2} z_{2}) \geq 2 \nonumber \\
(1 + \sigma_{1} z_{1})(1 + \sigma_{2} z_{2}) \geq 2
\end{eqnarray} 
are not compatible for the variables $z_{i} = \tanh \beta E_{i}$, $i = 1,2$.
Indeed, let $\epsilon_{1} \neq \sigma_{1}$,
i. e. $\epsilon_{1} \sigma_{1} = - 1$. Multiplying the inequalities 
(\ref{4.17}) we obtain the inequality
\begin{equation}
\label{4.18}
(1 - z_{1}^{2})(1 + \epsilon_{2} z_{2})(1 + \sigma_{2} z_{2}) \geq 4
\end{equation} 
Since $z_{i} = \tanh \beta E_{i}$, $\beta > 0$, $E_{i} \neq 0$, $i = 1,2$, 
then $0 < |z_{i}| < 1$ and
\begin{equation}
\label{4.19}
0 < 1 - z_{1}^{2} < 1, 0 < 1 + \epsilon_{2} z_{2} < 2,
0 < 1 + \sigma_{2} z_{2} < 2.
\end{equation}
The inequalities (\ref{4.18}) and (\ref{4.19}) are not compartible.
The case $\epsilon_{2} \neq \sigma_{2}$ is considered similarly.

Let us prove that for the variables
$z_{i} = \tanh \beta E_{i}$, $\beta > 0$, $E_{i} \neq 0$, $i = 1,2$, 
the first inequality (\ref{4.17}) implies $\epsilon_{i} z_{i} > 0$,
$i = 1,2$. Indeed, for the variables $z_{i}$ the inequality
$0 < 1 + \epsilon_{i} z_{i} < 2$ is valid. Hence the first inequality
(\ref{4.17}) implies $1 + \epsilon_{i} z_{i} > 1$ or $\epsilon_{i} z_{i} > 0$.
The lemma is proved.

\noindent {\bf Theorem 4.3} {\it For the partition function} (\ref{2.2})
{\it the following relation is valid}
\begin{eqnarray}
\label{4.20}
- \beta^{- 1} \lim_{M,N \rightarrow \infty} (MN)^{- 1}\ln Z(N,M) = \nonumber \\
- \beta^{- 1} \ln (2\cosh \beta E_{1}\cosh \beta E_{2}) -
1/2\beta^{- 1} (2\pi )^{- 2} \int_{0}^{2\pi} d\theta_{1} \int_{0}^{2\pi} 
d\theta_{2} \nonumber \\
\ln [(1 + z_{1}^{2})(1 + z_{2}^{2}) - 2z_{1}(1 - z_{2}^{2})\cos \theta_{1}
- 2z_{2}(1 - z_{1}^{2})\cos \theta_{2} ]
\end{eqnarray}
{\it where the variables}
$z_{i} = \tanh \beta E_{i}$, $\beta > 0$, $E_{i} \neq 0$, $i = 1,2$.

\noindent {\it Proof.} Let the numbers $N,M$ be odd. Then the homological
formula (\ref{2.42}) and the equalities (\ref{3.23}), (\ref{3.31}), 
(\ref{3.37}), (\ref{3.43}) imply
\begin{eqnarray}
\label{4.21}
Z(N,M) = (2\cosh \beta E_{1}\cosh \beta E_{2})^{MN} \times \nonumber \\
1/2[F_{11}(z_{1},z_{2})((1 + z_{1})(1 + z_{2}) - 2) -
F_{21}(z_{1},z_{2})((1 - z_{1})(1 + z_{2}) - 2) - \nonumber \\
- F_{31}(z_{1},z_{2})((1 + z_{1})(1 - z_{2}) - 2) - 
F_{41}(z_{1},z_{2})((1 - z_{1})(1 - z_{2}) - 2)].
\end{eqnarray}
Since the function $\cos x$ is monotone decreasing on the interval
$[0,\pi ]$, we have
\begin{eqnarray}
\label{4.22}
\cos (2\pi N^{- 1}j) < \cos (\pi N^{- 1}(2j - 1)), 1 \leq j \leq 
\frac{N - 1}{2}, \nonumber \\
\cos(\pi N^{- 1}(2(j + 1) - 1)) < \cos (2\pi N^{- 1}j),
0 \leq j \leq \frac{N - 3}{2}.
\end{eqnarray}
It follows from the relations (\ref{3.24}), (\ref{3.32}), (\ref{3.38}),
(\ref{3.44}) and the inequalities (\ref{4.22}) that for $z_{1} > 0$,
$z_{2} > 0$
\begin{eqnarray}
\label{4.23}
F_{21}(z_{1},z_{2}) < \nonumber \\
P_{21}(\frac{N + 1}{2} ,\frac{N + 1}{2} ;1,\frac{M - 1}{2} )
(P_{11}(0,0;1,\frac{M - 1}{2} ))^{- 1}
F_{11}(z_{1},z_{2})
\end{eqnarray}
\begin{eqnarray}
\label{4.24}
F_{21}(z_{1},z_{2}) > 
P_{21}(1,1;0,\frac{M - 1}{2} )
P_{21}(\frac{N + 1}{2}, \frac{N + 1}{2}; 1,\frac{M - 1}{2} ) \times 
\nonumber \\
(P_{11}(\frac{N - 1}{2} ,\frac{N - 1}{2} ;0,\frac{M - 1}{2} )
P_{11}(\frac{N - 1}{2} ,\frac{N - 1}{2} ;1,\frac{M - 1}{2} ))^{- 1}
F_{11}(z_{1},z_{2})
\end{eqnarray}
\begin{eqnarray}
\label{4.25}
F_{31}(z_{1},z_{2}) < \nonumber \\
P_{12}(1,\frac{N - 1}{2} ;\frac{M + 1}{2} ,\frac{M + 1}{2} )
(P_{11}(1,\frac{N - 1}{2} ;0,0))^{- 1}
F_{11}(z_{1},z_{2})
\end{eqnarray}
\begin{eqnarray}
\label{4.26}
F_{31}(z_{1},z_{2}) > 
P_{12}(0,\frac{N - 1}{2} ;1,1)
P_{12}(1, \frac{N - 1}{2}; \frac{M + 1}{2} ,\frac{M + 1}{2} ) \times 
\nonumber \\
(P_{11}(0,\frac{N - 1}{2} ;\frac{M - 1}{2} ,\frac{M - 1}{2} )
P_{11}(1,\frac{N - 1}{2} ;\frac{M - 1}{2} ,\frac{M - 1}{2} ))^{- 1}
F_{11}(z_{1},z_{2})
\end{eqnarray}
\begin{eqnarray}
\label{4.27}
F_{41}(z_{1},z_{2}) < 
P_{22}(1,\frac{N - 1}{2} ;\frac{M + 1}{2} ,\frac{M + 1}{2} )
P_{22}(\frac{N + 1}{2} , \frac{N + 1}{2}; 1,\frac{M - 1}{2} ) \times 
\nonumber \\
(P_{11}(1,\frac{N - 1}{2} ;0,0)
P_{11}(0,0;1,\frac{M - 1}{2} ))^{- 1}
F_{11}(z_{1},z_{2})
\end{eqnarray}
\begin{eqnarray}
\label{4.28}
F_{41}(z_{1},z_{2}) > 
P_{22}(1,\frac{N - 1}{2} ;1,1)
P_{22}(1,1;1,\frac{M - 1}{2} ) \times 
\nonumber \\
P_{22}(1,\frac{N - 1}{2} ;\frac{M + 1}{2} ,\frac{M + 1}{2} )
P_{22}(\frac{N + 1}{2} , \frac{N + 1}{2}; 1,\frac{M - 1}{2} ) \times 
\nonumber \\
(P_{11}(0,\frac{N - 3}{2} ;\frac{M - 1}{2} ,\frac{M - 1}{2} )
P_{11}(\frac{N - 1}{2} ,\frac{N - 1}{2} ;0,\frac{M - 3}{2} ))^{- 1}
\times \nonumber \\
(P_{11}(1,\frac{N - 1}{2} ;\frac{M - 1}{2} ,\frac{M - 1}{2} )
P_{11}(\frac{N - 1}{2} ,\frac{N - 1}{2} ;1,\frac{M - 1}{2} ))^{- 1}
F_{11}(z_{1},z_{2}).
\end{eqnarray}

The relation (\ref{4.21}) and the inequalities (\ref{4.23}), (\ref{4.25}),
(\ref{4.27}) imply for the domains (\ref{4.5}), (\ref{4.6}) and (\ref{4.13})
\begin{eqnarray}
\label{4.29}
2Z(N,M)(2\cosh \beta E_{1}\cosh \beta E_{2})^{- MN}(F_{11}(z_{1},z_{2}))^{- 1}
< \nonumber \\
((1 + z_{1})(1 + z_{2}) - 2) - \nonumber \\
- ((1 - z_{1})(1 + z_{2}) - 2)
P_{21}(\frac{N + 1}{2} ,\frac{N + 1}{2} ;1,\frac{M - 1}{2} )
(P_{11}(0,0;1,\frac{M - 1}{2} ))^{- 1} - \nonumber \\
- ((1 + z_{1})(1 - z_{2}) - 2)
P_{12}(1,\frac{N - 1}{2} ;\frac{M + 1}{2} ,\frac{M + 1}{2} )
(P_{11}(1,\frac{N - 1}{2} ;0,0))^{- 1} - \nonumber \\
- ((1 - z_{1})(1 - z_{2}) - 2)
P_{22}(1,\frac{N - 1}{2} ;\frac{M + 1}{2} ,\frac{M + 1}{2} )
P_{22}(\frac{N + 1}{2}, \frac{N + 1}{2} ;1,\frac{M - 1}{2} ) \times
\nonumber \\
(P_{11}(1,\frac{N - 1}{2} ;0,0)
P_{11}(0,0;1,\frac{M - 1}{2} ))^{- 1}.
\end{eqnarray}

By making use of the inequalities (\ref{4.22}) we have
\begin{eqnarray}
\label{4.30}
P_{11}(1,\frac{N - 1}{2} ;0,0)
(P_{22}(1,\frac{N - 1}{2} ;\frac{M + 1}{2} ,\frac{M + 1}{2} ))^{- 1} 
< \nonumber \\
((1 + z_{1}^{2})(1 + z_{2}^{2})  
+ 2z_{1}(1 - z_{2}^{2})\cos (\pi N^{- 1})
- 2z_{2}(1 - z_{1}^{2})) \times \nonumber \\
((1 + z_{1}^{2})(1 + z_{2}^{2})  
- 2z_{1}(1 - z_{2}^{2})\cos (\pi N^{- 1})
+ 2z_{2}(1 - z_{1}^{2}))^{- 1}
\end{eqnarray}
\begin{eqnarray}
\label{4.31}
P_{11}(0,0;1,\frac{M - 1}{2} )
(P_{22}(\frac{M + 1}{2} ,\frac{N + 1}{2} ;1,\frac{M - 1}{2} ))^{- 1} 
< \nonumber \\
((1 + z_{1}^{2})(1 + z_{2}^{2}) 
- 2z_{1}(1 - z_{2}^{2})
+ 2z_{2}(1 - z_{1}^{2})\cos (\pi M^{- 1})) \times \nonumber \\
((1 + z_{1}^{2})(1 + z_{2}^{2})  
+ 2z_{1}(1 - z_{2}^{2})
- 2z_{2}(1 - z_{1}^{2})\cos (\pi M^{- 1}))^{- 1}
\end{eqnarray}
\begin{eqnarray}
\label{4.32}
P_{21}(\frac{N + 1}{2} ,\frac{N + 1}{2} ;1,\frac{M - 1}{2} )
(P_{22}(\frac{N + 1}{2} ,\frac{N + 1}{2} ;1,\frac{M - 1}{2} ))^{- 1} 
< \nonumber \\
((1 + z_{1}^{2})(1 + z_{2}^{2})  
+ 2z_{1}(1 - z_{2}^{2})
+ 2z_{2}(1 - z_{1}^{2})\cos (\pi M^{- 1})) \times \nonumber \\
((1 + z_{1}^{2})(1 + z_{2}^{2})  
+ 2z_{1}(1 - z_{2}^{2})
- 2z_{2}(1 - z_{1}^{2})\cos (\pi M^{- 1}))^{- 1}
\end{eqnarray}
\begin{eqnarray}
\label{4.33}
P_{12}(1,\frac{N - 1}{2} ;\frac{M + 1}{2} ,\frac{M + 1}{2} )
(P_{22}(1,\frac{N - 1}{2} ;\frac{M + 1}{2} ,\frac{M + 1}{2} ))^{- 1} 
< \nonumber \\
((1 + z_{1}^{2})(1 + z_{2}^{2})  
+ 2z_{1}(1 - z_{2}^{2})\cos (\pi N^{- 1})
+ 2z_{2}(1 - z_{1}^{2})) \times \nonumber \\
((1 + z_{1}^{2})(1 + z_{2}^{2})  
- 2z_{1}(1 - z_{2}^{2})\cos (\pi N^{- 1})
+ 2z_{2}(1 - z_{1}^{2}))^{- 1}.
\end{eqnarray}
In view of Lemma 4.1 the right hand sides of the equalities (\ref{4.30}) -
(\ref{4.33}) are bounded.

By dividing the inequailty (\ref{4.29}) by

$$P_{22}(1,\frac{N - 1}{2} ;\frac{M + 1}{2} ,\frac{M + 1}{2} )
P_{22}(\frac{N + 1}{2}, \frac{N + 1}{2} ;1,\frac{M - 1}{2} ) \times $$

$${(P_{11}(1,\frac{N - 1}{2} ;0,0)P_{11}(0,0;1,\frac{M - 1}{2} ))^{- 1}}$$

and by using the inequalities (\ref{4.30}) - (\ref{4.33}) we have
\begin{eqnarray}
\label{4.34}
- \beta^{- 1} \lim_{M,N \rightarrow \infty} (MN)^{- 1}\ln Z(N,M) \geq
\nonumber \\
- \beta^{- 1} \ln (2\cosh \beta E_{1}\cosh \beta E_{2}) -
1/2\beta^{- 1} (2\pi )^{- 2} \int_{0}^{2\pi} d\theta_{1} \int_{0}^{2\pi} 
d\theta_{2} \nonumber \\
\ln [(1 + z_{1}^{2})(1 + z_{2}^{2}) - 2z_{1}(1 - z_{2}^{2})\cos \theta_{1}
- 2z_{2}(1 - z_{1}^{2})\cos \theta_{2} ]
\end{eqnarray}
for the domains (\ref{4.5}), (\ref{4.6}) and (\ref{4.13}).

The relation (\ref{4.21}) and the inequalities (\ref{4.24}), (\ref{4.26}),
(\ref{4.28}) imply for the domains (\ref{4.5}), (\ref{4.6}) and (\ref{4.13})
\begin{eqnarray}
\label{4.35}
2Z(N,M)(2\cosh \beta E_{1}\cosh \beta E_{2})^{- MN}(F_{11}(z_{1},z_{2}))^{- 1}
> \nonumber \\
((1 + z_{1})(1 + z_{2}) - 2) - \nonumber \\
- ((1 - z_{1})(1 + z_{2}) - 2)
P_{21}(1,1;0,\frac{M - 1}{2} )
P_{21}(\frac{N + 1}{2} ,\frac{N + 1}{2} ;1,\frac{M - 1}{2} ) \times 
\nonumber \\
(P_{11}(\frac{N - 1}{2} ,\frac{N - 1}{2} ;0,\frac{M - 1}{2} )
P_{11}(\frac{N - 1}{2} ,\frac{N - 1}{2} ;1,\frac{M - 1}{2} ))^{- 1} 
- \nonumber \\
- ((1 + z_{1})(1 - z_{2}) - 2)
P_{12}(0,\frac{N - 1}{2} ;1,1)
P_{12}(1,\frac{N - 1}{2} ;\frac{M + 1}{2} ,\frac{M + 1}{2} )
\times \nonumber \\
(P_{11}(0,\frac{N - 1}{2} ;\frac{M - 1}{2} ,\frac{M - 1}{2} )
P_{11}(1,\frac{N - 1}{2} ;\frac{M - 1}{2} ,\frac{M - 1}{2} ))^{- 1}
- \nonumber \\
- ((1 - z_{1})(1 - z_{2}) - 2)
P_{22}(1,\frac{N - 1}{2} ;1,1)
P_{22}(1,1;1,\frac{M - 1}{2} ) \times
\nonumber \\
P_{22}(1,\frac{N - 1}{2} ;\frac{M + 1}{2} ,\frac{M + 1}{2} )
P_{22}(\frac{N + 1}{2}, \frac{N + 1}{2} ;1,\frac{M - 1}{2} ) \times
\nonumber \\
(P_{11}(0,\frac{N - 3}{2} ;\frac{M - 1}{2} ,\frac{M - 1}{2} )
P_{11}(\frac{N - 1}{2} ,\frac{N - 1}{2} ;0,\frac{M - 3}{2} ))^{- 1} 
\times \nonumber \\
(P_{11}(1,\frac{N - 1}{2} ;\frac{M - 1}{2} ,\frac{M - 1}{2} )
P_{11}(\frac{N - 1}{2} ,\frac{N - 1}{2} ;1,\frac{M - 1}{2} ))^{- 1}.
\end{eqnarray}

By using the inequalities (\ref{4.22}) it is easy to obtain for the
domains (\ref{4.5}), (\ref{4.6}), (\ref{4.13}) for large numbers
$N,M$ the inequalities similar to the inequalities (\ref{4.30}) -
(\ref{4.33}) for the products

$$P_{21}(1,1;0,\frac{M - 1}{2} )
P_{21}(\frac{N + 1}{2} ,\frac{N + 1}{2} ;1,\frac{M - 1}{2} ) \times $$

$$(P_{11}(\frac{N - 1}{2} ,\frac{N - 1}{2} ;0,\frac{M - 1}{2} )
P_{11}(\frac{N - 1}{2} ,\frac{N - 1}{2} ;1,\frac{M - 1}{2} ))^{- 1}, $$

$$P_{12}(0,\frac{N - 1}{2} ;1,1)
P_{12}(1,\frac{N - 1}{2} ;\frac{M + 1}{2} ,\frac{M + 1}{2} ) \times $$

$$(P_{11}(0,\frac{N - 1}{2} ;\frac{M - 1}{2} ,\frac{M - 1}{2} )
P_{11}(1,\frac{N - 1}{2} ;\frac{M - 1}{2} ,\frac{M - 1}{2} ))^{- 1},$$

$$P_{22}(1,\frac{N - 1}{2} ;1,1)
P_{22}(1,1;1,\frac{M - 1}{2} ) \times $$

$$P_{22}(1,\frac{N - 1}{2} ;\frac{M + 1}{2} ,\frac{M + 1}{2} )
P_{22}(\frac{N + 1}{2}, \frac{N + 1}{2} ;1,\frac{M - 1}{2} ) \times $$

$$(P_{11}(0,\frac{N - 3}{2} ;\frac{M - 1}{2} ,\frac{M - 1}{2} )
P_{11}(\frac{N - 1}{2} ,\frac{N - 1}{2} ;0,\frac{M - 3}{2} ))^{- 1} \times $$

$$(P_{11}(1,\frac{N - 1}{2} ;\frac{M - 1}{2} ,\frac{M - 1}{2} )
P_{11}(\frac{N - 1}{2} ,\frac{N - 1}{2} ;1,\frac{M - 1}{2} ))^{- 1}.$$

Therefore the inequality (\ref{4.35}) implies the following inequality
\begin{eqnarray}
\label{4.36}
- \beta^{- 1} \lim_{M,N \rightarrow \infty} (MN)^{- 1}\ln Z(N,M) \leq
\nonumber \\
- \beta^{- 1} \ln (2\cosh \beta E_{1}\cosh \beta E_{2}) -
1/2\beta^{- 1} (2\pi )^{- 2} \int_{0}^{2\pi} d\theta_{1} \int_{0}^{2\pi} 
d\theta_{2} \nonumber \\
\ln [(1 + z_{1}^{2})(1 + z_{2}^{2}) - 2z_{1}(1 - z_{2}^{2})\cos \theta_{1}
- 2z_{2}(1 - z_{1}^{2})\cos \theta_{2} ]
\end{eqnarray}
for the domains (\ref{4.5}), (\ref{4.5}), (\ref{4.13}). The equality 
(\ref{4.20}) for the domains (\ref{4.5}), (\ref{4.6}), (\ref{4.13})
follows from the inequalities (\ref{4.34}), (\ref{4.36}).

Let the numbers $N,M$ be odd and $z_{1} < 0$, $z_{2} > 0$. The relations
(\ref{3.24}), (\ref{3.32}), (\ref{3.38}), (\ref{3.44}) and inequalities
(\ref{4.22}) imply
\begin{eqnarray}
\label{4.37}
F_{11}(z_{1},z_{2}) < \nonumber \\
P_{11}(0,0;1,\frac{M - 1}{2} )
(P_{21}(\frac{N + 1}{2} ,\frac{N + 1}{2} ;1,\frac{M - 1}{2} ))^{- 1}
F_{21}(z_{1},z_{2})
\end{eqnarray}
\begin{eqnarray}
\label{4.38}
F_{11}(z_{1},z_{2}) > 
P_{11}(\frac{N - 1}{2} ,\frac{N - 1}{2} ;1,\frac{M - 1}{2} )
P_{11}(\frac{N - 1}{2}, \frac{N - 1}{2}; 0,\frac{M - 1}{2} ) \times 
\nonumber \\
(P_{21}(\frac{N + 1}{2} ,\frac{N + 1}{2} ;1,\frac{M - 1}{2} )
P_{21}(1,1;0,\frac{M - 1}{2} ))^{- 1}
F_{21}(z_{1},z_{2})
\end{eqnarray}
\begin{eqnarray}
\label{4.39}
F_{31}(z_{1},z_{2}) < 
P_{12}(0,0;1,\frac{M - 1}{2} )
P_{12}(1,\frac{N - 1}{2} ;\frac{M + 1}{2} ,\frac{M + 1}{2} ) \times
\nonumber \\
(P_{21}(1,\frac{N - 1}{2} ;0,0)
P_{21}(\frac{N + 1}{2} ,\frac{N + 1}{2} ;1,\frac{M - 1}{2} ))^{- 1}
F_{21}(z_{1},z_{2})
\end{eqnarray}
\begin{eqnarray}
\label{4.40}
F_{31}(z_{1},z_{2}) > 
P_{21}(1,\frac{N - 1}{2} ;0,0)
P_{12}(1, \frac{N - 1}{2}; \frac{M + 1}{2} ,\frac{M + 1}{2} ) \times 
\nonumber \\
(P_{12}(\frac{N - 1}{2} ,\frac{N - 1}{2} ;1,\frac{M - 1}{2} ))^{2}
(P_{21}(1, \frac{N - 1}{2}; \frac{M - 1}{2} ,\frac{M - 1}{2} ))^{- 2} \times 
\nonumber \\
(P_{21}(1,1;0,\frac{M - 3}{2} )
P_{21}(\frac{N + 1}{2} ,\frac{N + 1}{2} ;1,\frac{M - 1}{2} ))^{- 1}
F_{21}(z_{1},z_{2})
\end{eqnarray}
\begin{eqnarray}
\label{4.41}
F_{41}(z_{1},z_{2}) < 
P_{22}(1,\frac{N - 1}{2} ;\frac{M + 1}{2} ,\frac{M + 1}{2} ) \times 
\nonumber \\
(P_{21}(1,\frac{N - 1}{2} ;0,0))^{- 1}
F_{21}(z_{1},z_{2})
\end{eqnarray}
\begin{eqnarray}
\label{4.42}
F_{41}(z_{1},z_{2}) > 
P_{21}(1,\frac{N + 1}{2} ;0,0)
P_{22}(1,\frac{N - 1}{2} ;\frac{M + 1}{2} ,\frac{M + 1}{2} ) \times 
\nonumber \\
(P_{21}(1,\frac{N + 1}{2} ;\frac{M - 1}{2} ,\frac{M - 1}{2} )
P_{21}(1,\frac{N - 1}{2} ;\frac{M - 1}{2} ,\frac{M - 1}{2} ))^{- 1}
F_{21}(z_{1},z_{2}).
\end{eqnarray}
By making use of the relation (\ref{4.21}) and the inequalities
(\ref{4.38}), (\ref{4.39}), (\ref{4.41}) it is possible to obtain
the inequality (\ref{4.34}) for the domains (\ref{4.7}), (\ref{4.8}),
(\ref{4.14}). Similarly the relation (\ref{4.21}) and the inequalities
(\ref{4.37}), (\ref{4.40}), (\ref{4.42}) imply the inequality (\ref{4.36})
for the domains (\ref{4.7}), (\ref{4.8}), (\ref{4.14}). The inequalities
(\ref{4.34}), (\ref{4.36}) for the domains (\ref{4.7}), (\ref{4.8}), 
(\ref{4.14}) imply the equality (\ref{4.20}) for these domains.

The proof of the equality (\ref{4.20}) for odd numbers $N,M$ and for the
domains (\ref{4.9}), (\ref{4.10}), (\ref{4.15}) is quite similar to the
proof of the equality for odd numbers $N,M$ and for the domains (\ref{4.7}),
(\ref{4.8}), (\ref{4.14}).

Let the numbers $N,M$ be odd and $z_{1} < 0$, $z_{2} < 0$. The relations
(\ref{3.24}), (\ref{3.32}), (\ref{3.38}), (\ref{3.44}) and the
inequalities (\ref{4.22}) imply
\begin{eqnarray}
\label{4.43}
F_{11}(z_{1},z_{2}) < 
P_{11}(1,\frac{N - 1}{2} ;0,0)P_{11}(0,0;1,\frac{M - 1}{2} )
\times \nonumber \\
(P_{22}(1,\frac{N - 1}{2} ;\frac{M + 1}{2} ,\frac{M + 1}{2} )
P_{22}(\frac{N + 1}{2} ,\frac{N + 1}{2} ;1,\frac{M - 1}{2} ))^{- 1}
F_{41}(z_{1},z_{2})
\end{eqnarray}
\begin{eqnarray}
\label{4.44}
F_{11}(z_{1},z_{2}) > 
P_{11}(1,\frac{N - 1}{2} ;\frac{M - 1}{2} ,\frac{M - 1}{2} )
P_{11}(0, \frac{N - 3}{2}; \frac{M - 1}{2} ,\frac{M - 1}{2} ) \times 
\nonumber \\
P_{11}(\frac{N - 1}{2} ,\frac{N - 1}{2} ;1,\frac{M - 1}{2} )
P_{11}(\frac{N - 1}{2}, \frac{N - 1}{2}; 0,\frac{M - 3}{2} ) \times 
\nonumber \\
(P_{22}(1,\frac{N - 1}{2} ;1,1)
P_{22}(1,\frac{N - 1}{2} ;\frac{M + 1}{2} ,\frac{M + 1}{2} ))^{- 1} 
\times \nonumber \\
(P_{22}(\frac{N + 1}{2} ,\frac{N + 1}{2} ;1,\frac{M - 1}{2} )
P_{22}(1,1;1,\frac{M - 1}{2} ))^{- 1}
F_{41}(z_{1},z_{2})
\end{eqnarray}
\begin{eqnarray}
\label{4.45}
F_{21}(z_{1},z_{2}) < \nonumber \\
P_{21}(1,\frac{N - 1}{2} ;0,0) 
(P_{22}(1,\frac{N - 1}{2} ;\frac{M + 1}{2} ,\frac{M + 1}{2} ))^{- 1}
F_{41}(z_{1},z_{2})
\end{eqnarray}
\begin{eqnarray}
\label{4.46}
F_{21}(z_{1},z_{2}) > 
P_{21}(1,\frac{N - 1}{2} ;\frac{M - 1}{2} ,\frac{M - 1}{2} )
P_{21}(1, \frac{N + 1}{2}; \frac{M - 1}{2} ,\frac{M - 1}{2} ) \times 
\nonumber \\
(P_{22}(1,\frac{N + 1}{2} ;1,1)
P_{22}(1, \frac{N - 1}{2}; \frac{M + 1}{2} ,\frac{M + 1}{2} ))^{- 1} 
F_{41}(z_{1},z_{2})
\end{eqnarray}
\begin{eqnarray}
\label{4.47}
F_{31}(z_{1},z_{2}) < \nonumber \\
P_{12}(0,0;1,\frac{M - 1}{2} ) 
(P_{22}(\frac{N + 1}{2} ,\frac{N + 1}{2} ;1,\frac{M - 1}{2} ))^{- 1}
F_{41}(z_{1},z_{2})
\end{eqnarray}
\begin{eqnarray}
\label{4.48}
F_{31}(z_{1},z_{2}) > 
P_{12}(\frac{N - 1}{2} ,\frac{N - 1}{2} ;1,\frac{M - 1}{2} )
P_{12}(\frac{N - 1}{2} ,\frac{N - 1}{2} ;1,\frac{M + 1}{2} ) \times 
\nonumber \\
(P_{22}(1,1;1,\frac{M + 1}{2} )
P_{22}(\frac{N + 1}{2} ,\frac{N + 1}{2} ;1,\frac{M - 1}{2} ))^{- 1}
F_{41}(z_{1},z_{2}).
\end{eqnarray}
The inequalities (\ref{4.43}) - (\ref{4.48}) imply the inequalities
(\ref{4.34}), (\ref{4.36}) and therefore the equality (\ref{4.20})
for the domains (\ref{4.11}), (\ref{4.12}), (\ref{4.16}). The equality
(\ref{4.20}) is proved for the odd numbers $N,M$.

The cases when $N$ is odd, $M$ is even; when $N$ is even, $M$ is odd and
when $N,M$ are even are analogous to the case when the numbres $N,M$ are
odd. The theorem is proved.

\end{document}